\documentclass[times]{MyArticle}

\usepackage[colorlinks,bookmarksopen,bookmarksnumbered,citecolor=blue,urlcolor=blue]{hyperref}

\usepackage[usenames,dvipsnames,svgnames,table]{xcolor}
\usepackage{tabularx}
\usepackage{algorithm}
\usepackage{algpseudocode}
\usepackage{amsthm}
\newtheorem*{remark}{Remark}
\usepackage{graphicx,subfig}

\begin{document}
\captionsetup[subfigure]{labelformat=empty}

\title{Volume and Mass Conservation in Lagrangian Meshfree Methods}

\author{Pratik Suchde\affil{1}\corrauth, 
Christian Leith\"auser\affil{2}, 
J\"org Kuhnert\affil{2}, 
St\'ephane P.A. Bordas\affil{1} } 

\address{\affilnum{1} Legato group, Faculty of Science, Technology and Medicine, University of Luxembourg, Luxembourg \break %
\affilnum{2}Fraunhofer ITWM, Germany.}

\corraddr{E-mail: pratik.suchde@gmail.com, pratik.suchde@uni.lu}

\begin{abstract}
Meshfree Lagrangian frameworks for free surface flow simulations do not conserve fluid volume. Meshfree particle methods like SPH are not mimetic, in the sense that discrete mass conservation does not imply discrete volume conservation. On the other hand, meshfree collocation methods typically do not use any notion of mass. As a result, they are neither mass conservative nor volume conservative at the discrete level. In this paper, we give an overview of various sources of conservation errors across different meshfree methods. The present work focuses on one specific issue: unreliable volume and mass definitions. We introduce the concept of representative masses and densities, which are essential for accurate post-processing especially in meshfree collocation methods. Using these, we introduce an artificial compression or expansion in the fluid to rectify errors in volume conservation. Numerical experiments show that the introduced frameworks significantly improve volume conservation behaviour, even for complex industrial test cases such as automotive water crossing.
\end{abstract}


\maketitle

\vspace{-6pt}


\section{Introduction}

Over the last few decades, meshfree or meshless methods have become a popular alternative to conventional mesh-based solution procedures. Their biggest advantage lie in avoiding the generation of a mesh to discretize the computational domain. Meshfree domain discretization procedures that produce a point cloud or a particle cloud are faster, and can be automated to a higher degree, even for complex domains \cite{Suchde2022_PCG}. Another major advantage of meshfree methods, especially when used in a Lagrangian framework, is the ease of capturing large deformations and displacements in the computational domain. In this context, the meshfree equivalent of remeshing is much cheaper and faster \cite{Suchde2019_MovingSurfaces}. These advantages of meshfree methods, naturally, come with their own challenges. One of the biggest drawbacks of meshfree methods is a lack of conservation. 

This lack of conservation can appear in two forms. The first is the lack of conservative properties inherent in derivative approximation procedures. This has been a topic of significant study in recent years \cite{Chiu2012, Suchde2017, Trask2020}, though several open questions still remain. The second issue of conservation, which has been largely ignored in meshfree literature, concerns the absence of mass and volume conservation. This becomes especially relevant in flow problems with free surfaces which are captured using Lagrangian frameworks. While the latter point is the topic of investigation of the present work, we explain the details of both notions of conservation in this paper. 

\begin{figure}
  \centering
  \includegraphics[width=0.9\textwidth]{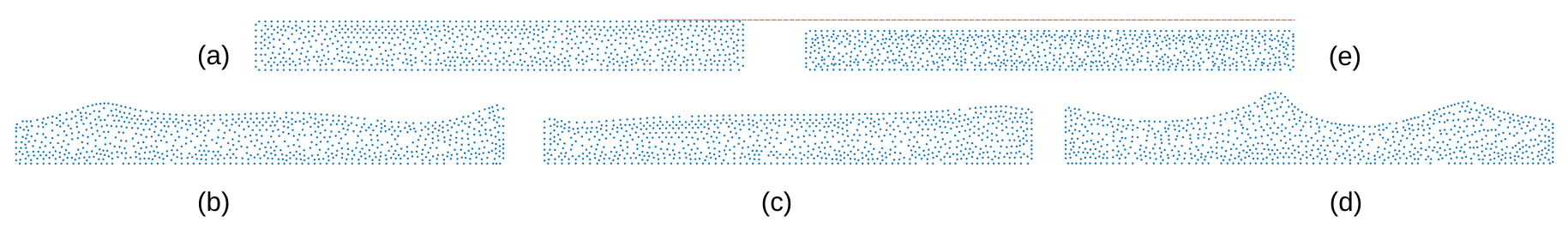}    
	\caption{Lack of volume conservation: Sloshing of an incompressible fluid with a particle method. The gravity vector is rotated to simulate sloshing, after which the fluid is brought back to rest. Counter-clockwise from top left. The top row shows the initial state (left) and the final state (right) of the fluid, while the bottom row shows intermediate states. The red dashed line indicates the level of the fluid at the initial state. The figures shows that the volume of the final state is lower than that present initially. Since the number of particles are constant, the total numerical mass is conserved. However, the fluid volume is not conserved.} 
  \label{Fig:VolCons_Illustration}%
\end{figure}

For the purpose of the present work, we consider meshfree methods for fluid flow divided into two categories. We refer the reader to review papers \cite{Belytschko1996, Nguyen2008, PUMbook} for an overview and classification of meshfree methods for structural simulations, which are not considered here. The first category of meshfree methods for fluid flow is particle-based meshfree methods such as Smoothed Particle Hydrodynamics (SPH) \cite{Gingold1977}, where each node of the domain discretization is a mass particle. The second category is meshfree collocation methods (for example, \cite{Jacquemin2020, Jacquemin2023}), where a node is only an approximation location, and does not inherently carry mass. The former mass-based nodes have the advantage of a simple notion of mass conservation. Since the mass of a particle is constant in time, and particles are neither added nor deleted during a simulation (barring inflows and outflows), the total mass contained in the system remains constant. As a result, these methods are typically said to be mass conservative. From a physical or continuum perspective, for incompressible flow, mass conservation and volume conservation are equivalent. However, this is not the case for SPH-type meshfree methods, where the fluid volume is not conserved, not even for incompressible flows with a constant density. In incompressible free-surface flow simulations, the geometric volume occupied by the computational domain often decreases as the simulation progresses, sometimes significantly so. An example of this is shown for a sloshing simulation in Figure~\ref{Fig:VolCons_Illustration}. On the other hand, for approximation node based meshfree collocation methods, since nodes do not carry mass, mass can only be defined as an auxiliary property.  As a result, there is no notion of mass or volume conservation in these methods.  

We note that this notion of volume conservation defects is also an issue for mesh-based methods for capturing free surface flow. Volume errors are observed in many mesh-based volume of fluid (VOF) approaches \cite{Arrufat2021}, as well as hybrid Eulerian mesh/Lagrangian particle methods \cite{Kugelstadt2019}. 

In mesh-based methods, the definition of an element or control volume prescribes a discrete volume packet. In contrast, for meshfree methods, there is no obvious notion of a discrete volume attached to a node. As a result, meshfree literature uses various different and unequivalent ways to define the volume of a point or particle. This compounds the issue of volume conservation errors, and also introduces errors in post-processing simulation results. In meshfree collocation type methods, a further issue in accurate post-processing is the lack of native mass definitions, which makes it cumbersome to track, for example, the transportation and accumulation of mass. 
A detailed example of this is presented later in Section~\ref{sec:WaterCrossing}.



The main contributions of this paper are four-fold: (1) We present a first of its kind survey of different conservation issues in meshfree methods for fluid flow. (2) We give an overview of different notions of defining discrete volumes for meshfree methods, and present their advantages and disadvantages. (3) We introduce the novel notion of representative masses and densities for meshfree methods, which simplify post-processing in collocation type methods. (4) Using the introduced representative masses and densities, we introduce an artificial point displacement to significantly reduce volume conservation errors. This is done based on the difference between the representative density and the actual (physical) density of the fluid. We note that the notion of artificial displacement has been widely adopted in the SPH community (for example, \cite{Lee2011, Lind2012, Pahar2017}). SPH literature uses artificial displacement to avoid particle clustering, which leads to numerical instabilities. In contrast, in the present work, we use a similar notion, but towards the goal of fluid volume conservation. While this work focuses on the mass and volume conservation in the context of collocation type meshfree methods, many of the ideas are also directly applicable to mass particle based meshfree methods. 




The paper is organized as follows. In Section~\ref{sec:Prelim} we introduce some further preliminaries about different types of meshfree methods, the notation used, and the governing equations of flow used in the present work. We then give a brief overview of different conservation issues across all meshfree methods in Section~\ref{sec:Conservation}. In Section~\ref{sec:Volume}, we present different methods for defining discrete volumes in meshfree methods. Section~\ref{sec:RepMass} then introduces the notion of representative masses, followed by the notion of representative densities in Section~\ref{sec:VolCorr}, which also introduces the volume correction algorithm. 
A series of numerical test cases are presented in Section~\ref{sec:Numerical} followed by a summary and conclusion in Section~\ref{sec:Conclusion}.


\section{Preliminaries}
\label{sec:Prelim}

In this section we introduce some preliminaries regarding the notation and governing equations used in the present work, along with a description of different types of meshfree methods. 

\subsection{Types of meshfree methods}
\label{sec:Meshfree}

As explained above, in the context of the current work, we divide Lagrangian meshfree methods for fluid flow into two broad classes, each with their own advantages and disadvantages. 
\begin{enumerate}
	\item Mass-particle based meshfree approaches. Here, each node is a mass particle. These types of methods include the popular Smoothed Particle Hydrodynamics (SPH) \cite{Gingold1977} and derived methods, such as Moving Particle Semi-explicit (MPS) \cite{Koshizuka1996}, Moving Particle Explicit (MPE \cite{Tayebi2015} or MPS \cite{Xiao2022}) and Discrete Droplet Method (DDM) \cite{Bharadwaj2022}, and also some hybrid mesh-meshfree methods like Material Point Methods (MPM) \cite{MPMbook}.
	\item Approximation-point based meshfree methods. Here, a node does not carry mass, and is only a collocation point where the governing PDEs are solved. While Radial Basis Function (RBF) based collocation methods have been widely used (for example, \cite{Depolli2022, Petras2018, Shankar2015}), Generalized Finite Difference Methods (GFDM) are more commonly used in Lagrangian approaches for fluid flow (for example, \cite{Drumm2008, Suchde2021}). 
\end{enumerate}

The main advantage of meshfree particle methods is that particles having mass directly implies a notion of mass conservation, which is not present in meshfree collocation approaches. On the other hand, the primary advantage of approximation point based meshfree methods is the ease of adding and merging/deleting points. This is very helpful to fix distortion as a result of movement. Points can easily be deleted or merged when they come too close, and points can be added when artificial ``holes" are created in the point cloud. This advantages also means that adaptive refinement is very straight forward in these methods. While SPH type methods are developing multi-resolution frameworks \cite{Ji2019, Zhang2021} with two or three discrete levels of refinement in the domain, collocation type meshfree methods have for long had the ability to achieve continuous resolution adaptivity \cite{Benito2003, Drumm2008, Gavete2003, Suchde2019_MovingSurfaces}. 


Without the ability to freely add and delete particles, particle distortion can not be fixed as easily in SPH type methods. Two common approaches are used for this: (i) redistributing particles based on a background mesh \cite{Obeidat2021, Obeidat2017, Obeidat2019}, or the more widely adopted (ii) artificially moving the particle cloud. 
For this, a non-physical or artificial movement of particles is performed to avoid particle clustering \cite{Lee2011, Lind2012, Pahar2017}. These approaches have been referred to as particle shifting. In the present work, we introduce the use of similar ideas to provide improved volume conservative behaviour. 

\begin{remark}
The topic of volume conservation has been widely discussed in recent SPH literature \cite{Jandaghian2021, Lyu2022, Pahar2016, Sun2019}. However, those discussions are in the context of particle shifting. Classical particle shifting methods introduced non-physical volume, while the volume conservation work in SPH aims to conserve volume during the particle shifting process. In contrast, the present work deals with volume conservation defects from other sources of the numerical scheme, and aims to perform an artificial point shifting to correct that. 
Another approach for volume conservation considered in the context of meshfree collocation methods is the so-called Position-Based Dynamics \cite{Basic2021, Basic2022, Basic2020}, which aims to keep the volume of each point constant as points move in a Lagrangian fashion. In contrast, here we aim to keep the global fluid volume constant.
\end{remark}

Another significant distinction between these two types of meshfree methods is the ease of enforcing boundary conditions. While a wide range of boundary conditions can be easily incorporated in meshfree collocation type methods, it is not straight forward to do the same in particle based meshfree methods \cite{Lind2020}.


\subsection{Notation}
\label{sec:Notation}

The bulk of this work uses an approximation-point based collocation meshfree method. We consider a computational domain $\Omega = \Omega(t)$, with boundary $\partial \Omega = \partial \Omega (t)$. The domain is discretized with a point cloud composed of $N=N(t)$ points, consisting of points both in the interior and on the boundary of the domain. 

For a point $i=1,2, ...N$, all approximations are done on a set of nearby nodes referred to as its neighbourhood or support $S_i$ consisting of $n_i$ nearby points, within a distance of $h=h(\vec{x}, t)$. $h$ denotes the so-called smoothing length or interaction radius, which governs not just the support size, but also the resolution of the point cloud. This is because inter-point distances are controlled to fixed multiples of $h$. The minimum distance between two points is controlled to be greater than $r_{min}h$, and maximum hole size in the point cloud is given by $r_{max}h$. This is enforced by merging particles that come too close, and adding them in holes where no points are present. For more details of these addition / deletion procedures, we refer to \cite{Farjoun2008, SeiboldThesis, Suchde2022_PCG, Suchde2019_MovingSurfaces}. 

The point clouds used in the present work are irregularly spaced, without any background grid or mesh. All points clouds are generated using an advancing front point generation method \cite{Suchde2022_PCG, Slak2019}. 

\subsection{Governing equations / model problem}
\label{sec:GoverningEquations}

The governing equations considered here are a unified framework for fluids and solids, consisting of the conservation of mass and momentum equations, used in a Lagrangian framework.
\begin{align}
	\frac{D \vec{x}}{Dt} &= \vec{v} \,,\\
	\frac{D \rho}{Dt}  &= - \rho \nabla \cdot \vec{v} \,,\\
	\frac{D \vec{v}}{Dt} &= \frac{1}{\rho} \nabla \cdot \mathbf{S} - \frac{1}{\rho} \nabla p  + \vec{g} \,, \label{Eq:Momentum}
\end{align}
for position $\vec{x}$, velocity $\vec{v}$, pressure $p$,  density $\rho$, gravity and other external forces $\vec{g}$, and a stress tensor $\mathbf{S}$ consisting of both viscous and solid terms. 
\begin{equation}
	\mathbf{S} = \mathbf{S}_{\text{visc}} + \mathbf{S}_{\text{solid}} \,,
\end{equation}
with
\begin{equation}
	\mathbf{S}_{\text{visc}} = \eta \left( 
	\left( \nabla \vec{v} \right)    +    \left( \nabla \vec{v} \right)^T 
	- \frac{2}{3} \nabla \cdot \vec{v} \,\mathbf{I}  \right) \,.
\end{equation}
The exact formulation of $\mathbf{S}_{\text{solid}}$ or of other material properties, such as $\rho$ or the viscosity $\eta$ is not very important in the present context. The ideas introduced in this paper could be used with a variety of material modelling frameworks. For simplicity, we use incompressible Newtonian flow with a constant density and viscosity, with $\mathbf{S}_{\text{solid}} = 0$ throughout this paper. If required, an energy conservation equation could also be used. While the ideas of this work can be easily applied to any governing equations, here we restrict the use to fluid flow. 


\section{Conservation and Meshfree Methods}
\label{sec:Conservation}

While we focus on one aspect of conservation, it is important to understand the other reasons of conservation defects to correctly interpret numerical results.  
In this section, we give a brief overview of different conservation related errors in meshfree methods for fluid flow. Each of these apply to both SPH-type and collocation meshfree methods. 
%
%
We note that these issues only cover Lagrangian meshfree methods for fluid flow. Eulerian weak form meshfree methods with stationary background meshes used in structural modelling are not covered here. 

\subsection{Conservation of fluxes}
\label{sec:FluxCons}

Consider a conservation law 
\begin{equation}
	\label{Eq:ConservationLaw}
	\frac{\partial\phi}{\partial t}+\nabla\cdot\mathbf{J}=0\,,
\end{equation}
with $\mathbf{J}=\mathbf{J}(\phi)$. For sufficiently smooth $\phi$ and $\mathbf{J}$, integrating Eq.\,\eqref{Eq:ConservationLaw} over the entire domain, and an application of the divergence theorem leads to the integral form of the conservation law,
\begin{equation}
	\label{Eq:EnergyConservation}
	\frac{d}{dt}\int_\Omega\phi\, dV=-\int_{\partial\Omega} \vec{n}\cdot \mathbf{J}\,dA \,.
\end{equation}

Due the local nature of meshfree methods and the absence of a global mesh discretizing the domain, most meshfree methods do not posses a discrete divergence theorem. As a result, at the discrete level, there is no connection between the differential law Eq.\,\eqref{Eq:ConservationLaw} and the integral form Eq.\,\eqref{Eq:EnergyConservation}. No matter how accurately the differential form is being solved at each node, there is no discrete equivalent of the global (integral) conservation law. 

Due to the absence of a mesh, a definition of flux is not natural to meshfree discretizations. To introduce global conservation in meshfree methods, 
\cite{Diyankov2008, Chiu2012, Trask2020} introduce different algebraic notions of fluxes to obtain a discrete divergence theorem. However, each of these methods require a global computation either of the differential operators \cite{Chiu2012} or of the fluxes \cite{Trask2020}. As a result, they are only used as a pre-processing step for meshfree simulations on fixed point clouds, and can be quite expensive for moving point clouds. To overcome this, \cite{Suchde2017} introduced a local definition of fluxes that is much quicker to compute, but only provides approximate conservation. For moving Lagrangian meshfree frameworks, achieving global conservation without expensive global computations remains an open problem. 

\subsection{Volume conservation}

As discussed earlier and illustrated in Figure~\ref{Fig:VolCons_Illustration}, the lack of volume conservation of the discrete domain is another source of numerical inaccuracy. This topic is the main focus of the present work. 

\subsubsection{Lagrangian motion}~\\
One of the biggest source of volume defects in Lagrangian methods is inaccurate advection. Lagrangian advection does not introduce diffusion, as is typically the case in Eulerian advection. It is thus often a preferred choice for capturing free surface flow. However, inaccurate advection for flows with free surfaces introduces volume conservation errors. 

Most Lagrangian meshfree methods consider a Lagrangian movement step by updating node positions using a first order accurate method, by treating the fluid velocity as constant between time steps. This has been shown to be extremely inaccurate, and higher order Lagrangian movement has been shown to significantly reduce volume conservation defects \cite{Suchde2018_PCM}. In our present work, we follow a higher order movement approach introduced in \cite{Suchde2018_PCM}, to reduce the influence of inaccurate Lagrangian motion on volume conservation errors. It is important to note here that this issue is relevant in all Lagrangian, semi-Lagrangian and ALE frameworks, covering meshfree, mesh-based, and hybrid methods. However, this aspect has been widely ignored in literature. 

The topic of volume conservation as a result of Lagrangian motion is usually mentioned in the context of the distortion of point or particle clouds. As described in the remark of Section~\ref{sec:Meshfree}, volume conservation in SPH generally refers to preventing volume changes in the artificial particle shifting step, and not overall. While in meshfree collocation methods, local volume conservation is enforced to conserve individual particle volumes \cite{Basic2020}.

\subsubsection{Mass and volume definitions}~\\
An important factor in the lack of volume conservation in Lagrangian meshfree methods is that a volume definition is not native to meshfree methods. This topic will be discussed in more detail in Section~\ref{sec:Volume}. 




\subsubsection{Other modelling and numerical inaccuracies}~\\
Volume conservation errors could also arise due to various other numerical inaccuracies. These could be due to the design of the complete numerical scheme (for instance, using a weakly compressible scheme to approximate incompressible equations), the tolerance of the iterative solvers, the (derivative) approximation error, or some other numerical source of error. 


Our present work combines issues relating too all these sources of volume errors. 
We introduce a volume correction scheme to address volume inaccuracies that could have arisen due to any of the above mentioned reasons.  


\section{Volume definition}
\label{sec:Volume}

Unlike the case of meshes, volume definitions are not native to meshfree methods. Before taking a look at the different approaches used in literature, we first define the concept of a consistent volume definition for a meshfree node cloud. A node here could be either a mass-carrying particle or an approximation point. For a volume definition to be consistent, the sum of volumes of all meshfree nodes in a domain should represent the physical volume occupied by the computational domain. For example, if a cube of side $l$ is discretized by meshfree nodes, the sum of volumes of all nodes should approximate the volume of the cube, $l^3$. Such a consistent volume definition is needed for accurate post-processing of simulation results. 

\begin{remark}
In the Lagrangian frameworks considered here, the node positions change in each time step. Thus, the volume of each node must be recomputed at every time step. 
\end{remark}

\begin{figure}
  \centering
  \subfloat[(a) Spherical volume]{\includegraphics[width=0.32\textwidth]{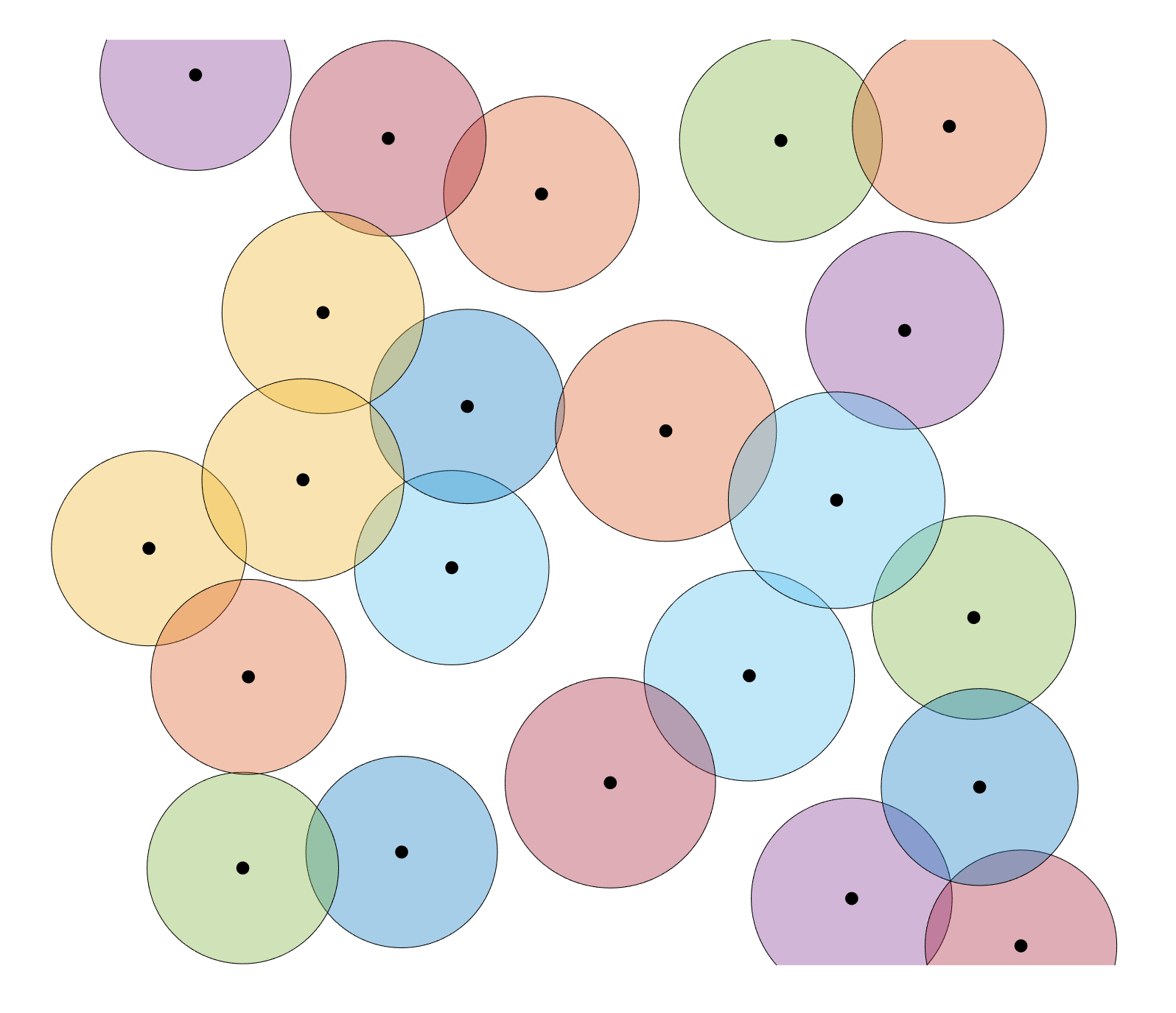}}
  \hfill
  \subfloat[(b) Global Voronoi decomposition]{\includegraphics[width=0.32\textwidth]{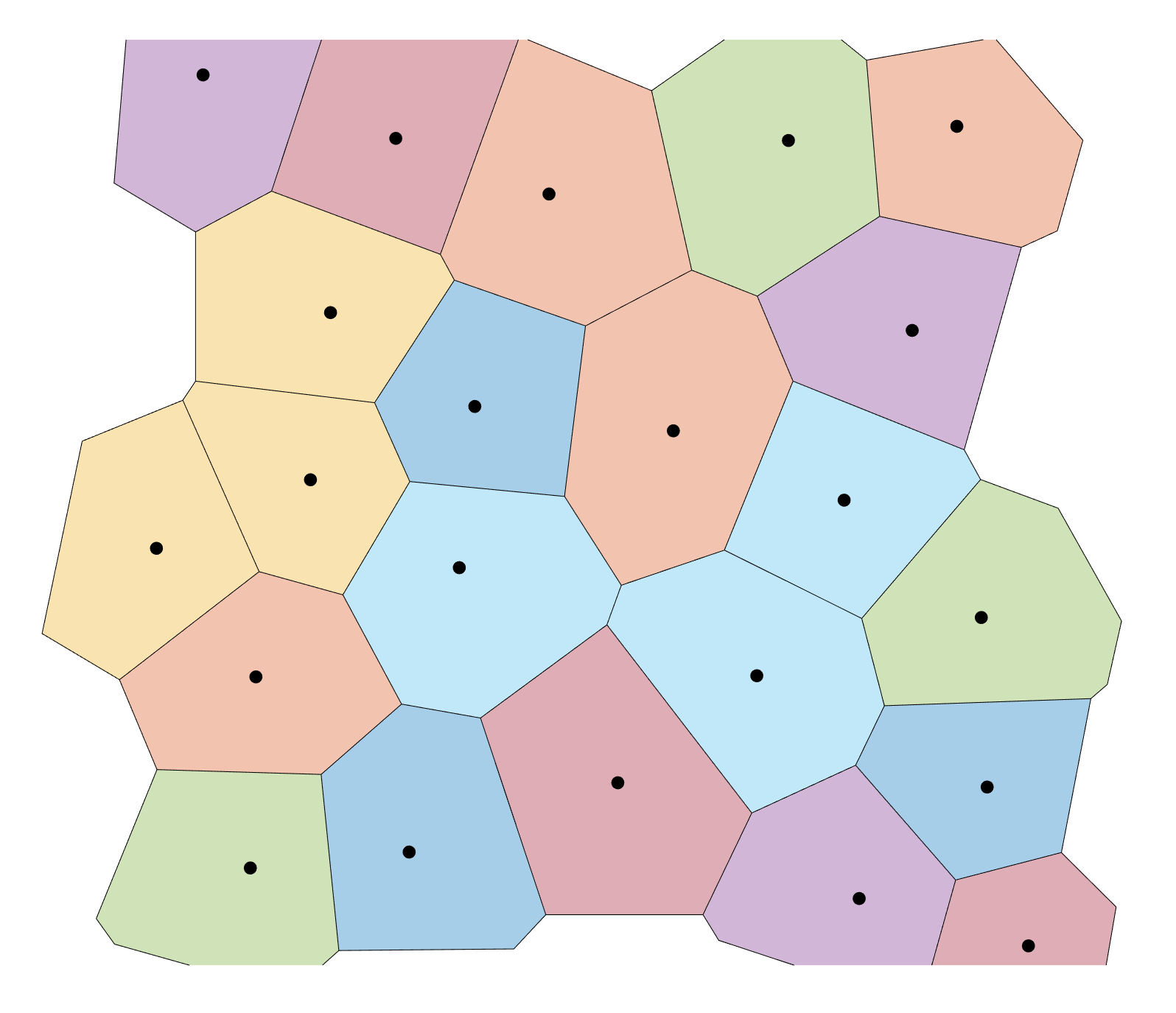}} 
  \hfill
  \subfloat[(c) Local Voronoi decomposition (with partial overlapping volumes)]{\includegraphics[width=0.32\textwidth]{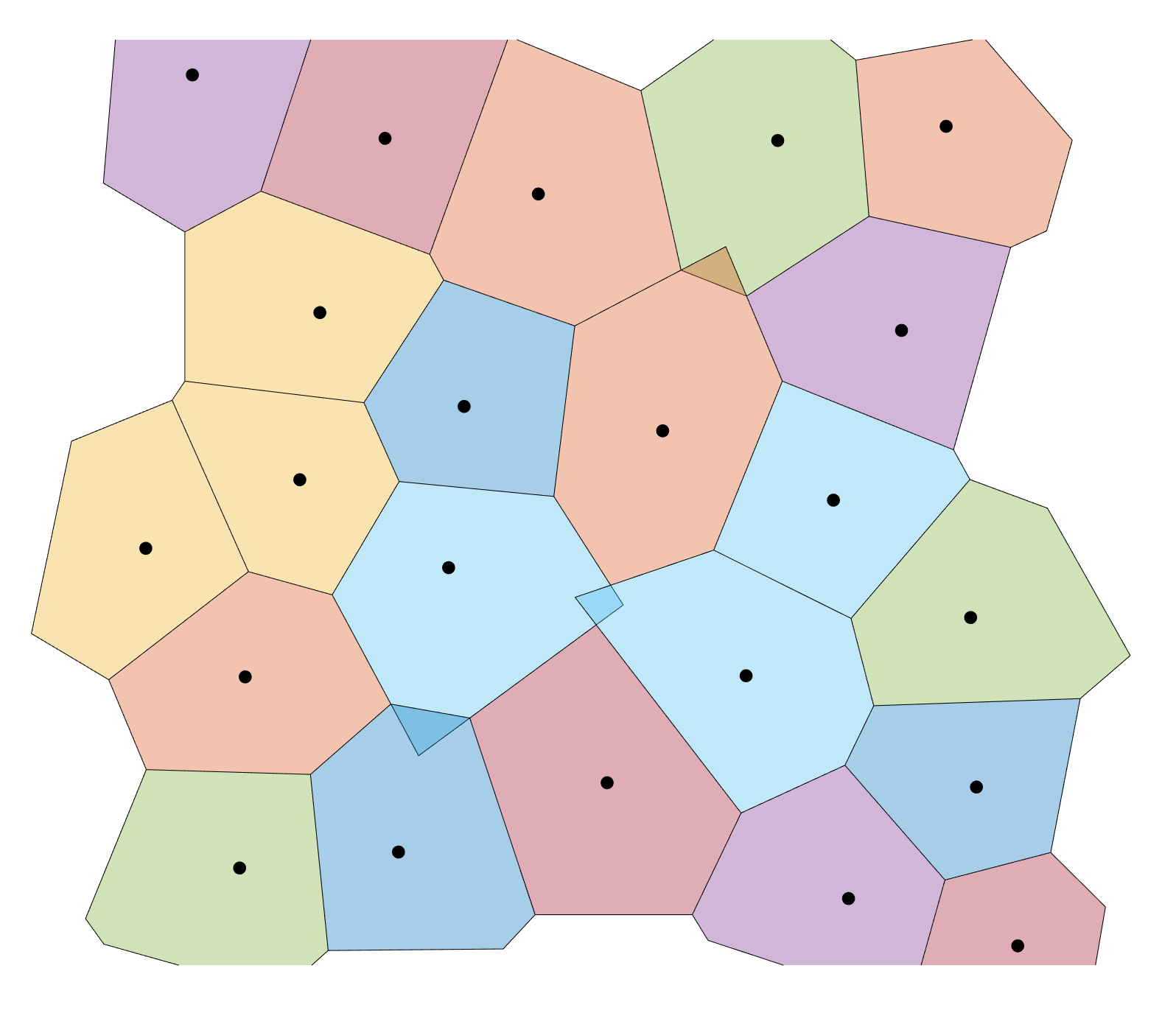}}
	\caption{Different volume definition of point cloud based meshfree methods.} 
  \label{Fig:VolumeDefinitions}%
\end{figure}

In meshfree literature, several different approaches have been considered to prescribe a notion of a volume to each meshfree point or particle. Both geometric and algebraic notions of volume have been proposed. A few commonly used examples are considered below:

\begin{itemize}
	\item The simplest approach is to associate a node with the volume of a sphere of a specified diameter. This could be based on the neighbourhood size or kernel width, or any other prescribed diameter. This method is not consistent due to either large overlapping volumes, or unaccounted regions in the middle of the domain, as is evident from Figure~\ref{Fig:VolumeDefinitions}.
	\item Another way to prescribe volumes to each point or particle is to construct a tessellation on the point cloud. Occasionally, for example \cite{Iske2007}, meshfree literature has used the notion of a globally constructed Voronoi tessellation of a point cloud. While this method is consistent, it can prove to be very expensive, especially when it needs to be reconstructed at every time step. 
	\item Another possible way to define volumes is through local tessellations. For each point $i$, its volume $V_i$ is given by the volume of the Voronoi cell locally defined on the support $S_i$. The locally defined Voronoi cells need not stitch together to form a global tessellation of the domain, as can be seen in Figure~\ref{Fig:VolumeDefinitions}. However, the amount of overlap is quite small \cite{Suchde2017} 
    and thus this method is consistent. It is also much faster than computing a global mesh on the set of points at every time step. We note here that actual local Voronoi tessellation need not be computed, as the shape of the elements and their faces are not needed. Figure~\ref{Fig:VolumeDefinitions} only shows the shapes for visualization. In reality \cite{Suchde2018_Thesis}, only the nearest neighbours in a Delaunay sense need to be identified locally, with the resultant simplex volumes giving the volume for the point in question. 
    \item In most mass-particle meshfree methods, each particle carries a mass and a density. This results in the obvious notion of volume as $V_i = \frac{m_i}{\rho_i}$, for mass $m$ and density $\rho$ \cite{Lind2020, Sigalotti2021}. During the initial particle generation process, masses are distributed to particles in a manner so as to make this method consistent. However, this method does not remain consistent once particles start moving. Other notions of volume include different algebraic manipulations of the mass and density \cite{Senz2022, Saitoh2013}. 
	\item Several other algebraic notions of volume have been proposed. These include virtual notions of volume to enforce a discrete divergence theorem, as explained in \ref{sec:FluxCons}. These methods typically solve global constrained optimisation problem to enforce an algebraic constraint, while maintaining a consistent volume definition in the whole domain \cite{Chiu2012}. However, due to the optimisation process, the volumes are no longer consistent on sub-domains. 
    \item Another algebraic approach is to use partition of unity functions. A differential volume at any arbitrary location is partitioned to the nearest nodes using a weighted kernel function with a normalization to ensure a partition of unity. This results in a Voronoi tessellation with smoothed edges \cite{Hopkins2015}. 
\end{itemize}





\begin{remark}
    Analogous to the definition of volume of each meshfree node is the question of prescribing an area for each boundary node. Similar approaches to the volume definition can be used here, though this topic has not been widely discussed.
\end{remark}



An important special case of prescribing volumes is for nodes with very few neighbours, which often occurs in violent free surface flows. Section~\ref{sec:WaterCrossing} is an example of such a flow pattern. When a node does not have sufficient neighbours, several of the consistent methods for defining a volume element explained above can not be used. 
In the present work, we use local Voronoi tessellations for defining volumes of each collocation point, as described in \cite{Suchde2018_Thesis, Suchde2017}. If the number of neighbours are insufficient to compute a local Voronoi tessellation, we use a spherical volume such that the volume of the point closely matches its volume before it became ``isolated".


\section{Representative Masses}
\label{sec:RepMass}

We now introduce the notion of representative mass for points in a meshfree collocation method. As a stand alone method, this does not affect the numerical scheme in any way, and is only a very useful post-processing tool. The need for the representative masses for accurate post-processing is highlighted in the numerical example in Section~\ref{sec:WaterCrossing}. This framework will also be the base on which the volume correction algorithm (see Section~\ref{sec:VolCorr} ) will be built. 

Consider a point $i$ with neighbourhood $S_i$, density $\rho_i$, and notional volumes $V_i$ computed as explained in Section~\ref{sec:Volume}. The representative mass of point $i$, denoted by $\hat{m}_i$, at the first time step is given simply by the physical density times the volume
\begin{equation}
    \label{Eq:RepMassInitial}
	\hat{m}_i (t=0) = \rho_i V_i \,.
\end{equation}

We note that unlike the case of mass particles in SPH-type methods, these representative masses can be shifted from one point to another when needed. This becomes essential when points are added or deleted for adaptive refinement, or to maintain a quasi-regularity of the point cloud. The procedures for the addition and deletion of points has been well established in meshfree collocation literature (for example, \cite{Drumm2008, Suchde2019_MovingSurfaces}), and will not be covered in the present work. The following subsections introduce a procedure to modify the representative masses as points are advected, added or deleted. 

\subsection{Mass adaptation}
\label{sec:MassAdaptation}

We first establish a general adaptation procedure for redistribution of representative masses when the point cloud changes. Special cases of this procedure will be used when points are added or deleted. The same procedure will also be used for a ``smoothing" of representative masses. If this smoothing procedure is not used, we empirically observe a high mass collection for a few points near where a lot of point adaption is needed, for example, near moving wall boundaries. 

For a point $i$ with representative mass $\hat{m}_i$ computed at the previous time step, the change in representative mass in the next time step is denoted by $\Delta \hat{m}_i$. We use the convention where a positive sign $\Delta \hat{m}_i > 0$ means that point $i$ gains mass. For the mass adaptation procedure, we first establish a set of conditions or rules that need to be fulfilled. 

\begin{enumerate}
    \item \label{item:MassCons} \emph{Total mass is constant}. Barring inflows and outflows (treated separately in Section~\ref{sec:InOut} ), no mass should be created or destroyed during the mass adaptation process. Thus, we have
    \begin{equation}
	   \label{Eq:MassConser}
	   \sum_{i = 1}^N \Delta \hat{m}_i = 0 \,,
    \end{equation}
    where $N = N(t)$ is the total number of points in the domain. Thus, any mass being added to a point must be removed from another, and vice-versa. For this, we introduce the notation $\Delta \hat{m}_{j \rightarrow i}$ to denote mass being transferred from point $j$ to point $i$. Thus, the change of representative mass of point $i$ can be given by
    \begin{equation}
        \label{Eq:Mass_GlobalSum}
        \Delta \hat{m}_i = \sum_{\substack{j=1 \\ j \neq i} }^N \Delta \hat{m}_{j \rightarrow i} \,.
    \end{equation}
    This uses the same convention as before, $\Delta \hat{m}_{j \rightarrow i} > 0$ implies point $j$ is losing mass, and point $i$ is gaining mass.
    \item \label{item:Local} \emph{Mass adaptation must be local}. This procedure has to be entirely local, in the sense that a mass redistribution from points $j$ to $i$ should only happen if $j$ is in the neighbourhood of $i$. Thus, 
    \begin{equation}
        \Delta \hat{m}_{j \rightarrow i} = 0 \,, \text{  if } j \notin S_i \,.
    \end{equation}
    Thus, Eq.\,\eqref{Eq:Mass_GlobalSum} reduces to
    \begin{equation}
        \label{Eq:Mass_LocalSum}
        \Delta \hat{m}_i = \sum_{\substack{j \in S_i \\ j \neq i} } \Delta \hat{m}_{j \rightarrow i} \,.
    \end{equation}
    \item \label{item:Dist} \emph{Distance dependence of mass adaptation}. Within the support domain of a point $i$, the mass transfer from a neighbouring point $j$ to $i$ should decrease as the distance between the points increases. 
    \item \label{item:Mass} \emph{Neighbour mass dependence of mass adaptation}. In the support domain of a point $i$, for equidistant neighbours $j_i$ and $j_2$, 
    \begin{equation}
        \hat{m}_{j_1} > \hat{m}_{j_2}  \implies \Delta \hat{m}_{j_1 \rightarrow i} \geq \Delta \hat{m}_{j_2 \rightarrow i} \,.
    \end{equation}
    More mass should be transferred from a neighbour with more mass, all other conditions being equal. This prevents mass clumping in certain points. 
    \item \label{item:LocalComp} \emph{Computations must be local}. The computation of the redistribution should be a local procedure. Since points move in a Lagrangian framework in each time step, and addition/deletion of points also has to be done in each time step, the mass adaptation procedure is run in each time step. Avoiding a global calculation of the representative mass redistribution can significantly speed up the redistribution algorithm. 
\end{enumerate}
Under theses rules defined above, the updated representative mass of each point should be modified to satisfy, if possible, 
\begin{equation}
	\label{Eq:Mv3}
	\hat{m}_i + \sum_{\substack{j \in S_i \\ j \neq i} } \Delta \hat{m}_{j \rightarrow i} = \hat{m}_i^{\text{target}} \,,
\end{equation}
where $\hat{m}_i^{\text{target}}$ is the target mass desired. For most cases, we have 
\begin{equation}
    \label{Eq:mTarget}
    \hat{m}_i^{\text{target}} =  \rho_i V_i\,,
\end{equation}
which relates the updated representative mass to the physical density and volume, as done during the initialization step, Eq.\,\eqref{Eq:RepMassInitial}. Outflow boundaries are an exception to this target mass prescription, and their treatment will be described in Section~\ref{sec:InOut}. For the introduced representative mass redistribution task, the following observations can be made
\begin{itemize}
    \item There is no unique solution. To enforce uniqueness, an additional optimality constraint would need to be added, for example, to minimize the representative mass changes, $\text{min} \sum_{i = 1}^N | \Delta \hat{m}_i |$. 
    \item The mass of a point $i$ can only change when the mass of at least one of its neighbours $j$ is also changing. Thus, determining any solution to this problem, with or without the added optimality constraint, would require a global procedure to determine the mass changes. However, this violates rule \ref{item:LocalComp}. 
\end{itemize}

To prevent the need for global computations, we modify the redistribution procedure to only approximate Eq.\,\eqref{Eq:Mv3}. In a first step, for each point $i$, local mass changes $\Delta \hat{m}_{j \rightarrow i}$ are computed to satisfy Eq.\,\eqref{Eq:Mv3} exactly. These mass changes are only computed and stored in this step, without determining the resultant new masses. Once the local mass changes for each point have been determined, they are communicated to neighbouring points across MPI processes. Finally, the sum of changes is applied to each point. Thus, a point $i$ will have a change in representative mass due the redistribution Eq.\,\eqref{Eq:Mv3} computed at point $i$ itself, and also for Eq.\,\eqref{Eq:Mv3} computed at each neighbouring point of $i$. For example, the net change of mass of point $i$ due to its neighbour $j$ is given by
\begin{equation}
    \Delta \hat{m}_{ij} = \Delta \hat{m}_{j \rightarrow i} - \Delta \hat{m}_{i \rightarrow j} \,,
\end{equation}
where $\Delta \hat{m}_{j \rightarrow i}$ is computed in the local computation of point $i$, and $\Delta \hat{m}_{i \rightarrow j}$ is computed in the local computation of point $j$. The entire procedure is also explained in the algorithm in Section~\ref{sec:RepMassAlgo}.

For the local redistribution process at each point $i$, we need to prescribe the form of  $\Delta \hat{m}_{j \rightarrow i}$. 
%
%
%
%
%
%
%
%
For this, we introduce the following ansatz 
\begin{equation}
	\label{Eq:DeltaM}
	\Delta \hat{m}_{j \rightarrow i} = \zeta_i K_{ij} \hat{m}_j  \,,
\end{equation}
with transfer coefficients $K_{ij}$ and a mass transfer function $\zeta_i$. By definition,  rule \ref{item:Mass} is satisfied. To satisfy the distance dependence rule \ref{item:Dist}, the transfer coefficients are based on the distance between the nearby points
\begin{equation}
	\label{Eq:Kij}
    K_{ij} =  \exp(-\alpha_K r_{ij}^2) \,,
\end{equation}
with a kernel width $\alpha_K = 2.0$, and where $r_{ij}$ is the relative distance between points $i$ and $j$
\begin{equation}
	r_{ij}^2 = \frac{ \|\vec{x}_j - \vec{x}_i \|^2 }{ \frac{h_i^2 + h_j^2}{2} } \,.
\end{equation}
Now, from Eqs.\,\eqref{Eq:Mv3}, \eqref{Eq:mTarget} and \eqref{Eq:DeltaM}, we get the mass transfer function
\begin{equation}
	\label{Eq:Zeta}
	\zeta_i = \frac{ \rho_i V_i - \hat{m}_i }{ \sum_{\substack{j \in S_i \\ j \neq i} } K_{ij}\hat{m}_j } \,.
\end{equation}
First, $\zeta_i$ is computed and stored for each point. After that, the representative masses are adapted for each point based on its own mass transfer condition as well as its neighbours' mass transfer conditions. 


\subsection{Addition of points}
\label{sec:Add}

As explained above, collocation points are added to the domain to prevent artificial hole formation due to the Lagrangian movement, and when adaptive refinement is desired. A representative mass is assigned to a newly added point $i$ by running the mass adaptation procedure explained in Section~\ref{sec:MassAdaptation} with $\hat{m}_i = 0$, indicating that the point had no mass before the redistribution procedure. The mass transfer function is then reduced to
\begin{equation}
	\zeta_i = \frac{ \rho_i V_i }{ \sum_{\substack{j \in S_i \\ j \neq i} } K_{ij}\hat{m}_j } \,.
\end{equation}
where $S_i$ is the new support of the newly added particle. In addition, we also need to set the transfer coefficient $K_{ij} = 0$ if point $j$ is also a newly added point in that time step. 


\subsection{Deletion / Merging of points}
\label{sec:Delete}

When two points come too close to each other, one of them is deleted, and the other is moved to the mid-point of the original two points. To adjust the representative masses, first all the relevant points are flagged for deletion. After that, the masses are redistributed, followed by the actual deletion of flagged points. If point $i$ is being deleted, or is being made inactive such that it no longer takes part in the approximations, we carry out the mass adaptation procedure laid out in Section~\ref{sec:MassAdaptation} with $V_i = 0$, indicating that the deleted point should have no volume. Thus, the mass transfer function given in Eq.\,\eqref{Eq:Zeta} becomes
\begin{equation}
	\zeta_i = \frac{ \hat{m}_i }{ \sum_{\substack{j \in S_i \\ j \neq i} } K_{ij}\hat{m}_j } \,.
\end{equation}
Additionally, if point $j$ is also flagged for deletion, we set the transfer coefficient $K_{ij}=0$, indicating that no mass is transferred between the two points. 

\subsection{Smoothing procedure}

To prevent an artificial accumulation of representative masses in a region, and to ensure that the representative masses are never negative, we run the general adaptation procedure established in Section~\ref{sec:MassAdaptation} for each point in the domain as a sort of smoothing of the masses. 

\subsection{Inflow and outflow boundaries}
\label{sec:InOut}

To maintain a notion of quasi-regularity at inflow and outflow boundaries, points on inflow and outflow boundaries are kept fixed in an Eulerian framework. The Lagrangian framework is only used for interior points, free surface points, slip boundary points, and interface points between two phases. Inflow points inject new points into the domain along the inward pointing normal. The new representative mass of these points is determined in the same as that for other newly added points, see Section~\ref{sec:Add}. 

For a point $i$ on an outflow boundary or a porous medium boundary, the target representative mass in Eq.\,\eqref{Eq:Mv3} is computed by considering the mass flux across the boundary
\begin{equation}
    \label{Eq:Outflow}
	\frac{d \hat{m}_i}{dt} \approx \frac{\hat{m}_i^{\text{target}} - \hat{m}_i}{\Delta t} = - \rho_i \left( \vec{v}_i \cdot \vec{n}_i \right) A_i \,,
\end{equation}
where $\vec{n}_i$ is the outward facing unit normal, and $A_i$ is the area occupied by the boundary point, computed with local tessellations in a similar manner to that done for volumes, as explained in Section~\ref{sec:Volume}. 

\subsection{Representative mass summary} 
\label{sec:RepMassAlgo}

The procedure for using the introduced representative mass as a post-processing tool in a meshfree collocation flow solver is summarized in Algorithm~\ref{alg:RepMass}, where the italicized steps indicate those not done in a typical flow solver. 

\begin{algorithm}
    \caption{Representative masses for post-processing} \label{alg:RepMass}
    \begin{algorithmic}[1]
        \State \emph{Initialize representative masses} (Eq.\,\eqref{Eq:RepMassInitial})
        \While{Time-stepping loop}
            \State Move point cloud in a Lagrangian sense \cite{Suchde2018_PCM}
            \State Update neighbour tree
            \State Add new points, flag points to be deleted
            \State Volume computation (Section~\ref{sec:Volume})
            \State \emph{Update representative masses for addition, deletion} (Sections~\ref{sec:Add}, \ref{sec:Delete})
            \State Delete points flagged for deletion
            \State \emph{Smooth representative masses} (Section~\ref{sec:MassAdaptation})
            \State Flow solver: velocity, pressure solve, stress tensor update if applicable
            \State Post-processing calculations
        \EndWhile
    \end{algorithmic}
\end{algorithm}


\section{Volume Correction Algorithm}
\label{sec:VolCorr}

Using the notion of representative mass introduced in Section~\ref{sec:RepMass}, we now introduce the volume correction mechanism. 

\subsection{Representative Density}
\label{sec:RepDens}
As a preliminary step for volume correction, we introduce a notion of representative density based on the updated representative masses as
%
\begin{equation}
\label{Eq:RepDens}
	\hat{\rho}_i = \frac{ \sum_{j \in S_i} \hat{m}_j }{\sum_{j \in S_i} V_j } \,.
\end{equation}
Note here that $i \in S_i$. We emphasize the notation difference: $\rho$ denotes the physical density, while $\hat{\rho}$ denotes the representative density. 

\subsection{Volume correction}
\label{sec:VolCorrSubSec}

Now we introduce the novel volume correction mechanism based on the relative difference between the representative density and the actual physical density. For each point $i$, a local correction is introduced which resembles an artificial velocity divergence. 
\begin{equation}
\label{Eq:DivVcorr}
  \left( \nabla \cdot \vec{v} \right) ^{\text{artificial}}_i = 
	 \frac{1}{\Delta t}
	 \text{min} \left\{ \frac{ \hat{\rho}_i - \rho_i }{\rho_i}, \gamma \right\}	  \,,
\end{equation}
where $\left( \nabla \cdot \vec{v} \right) ^{\text{artificial}}_i$ is the artificial or correction velocity divergence at point $i$, and $\gamma$ is a user defined parameter signifying the maximum allowed correction per time-step. Eq.\,\eqref{Eq:DivVcorr} prescribes and additional compression or expansion rate of the point cloud to artificially generate or destroy volume. 

The artificial velocity divergence introduced in Eq.\,\eqref{Eq:DivVcorr} can be used to correct the fluid volume in one of two ways. The first is to include it within the main numerical scheme. This can be done by adding the term in the right hand side of the pressure Poisson equation, which is straight forward irrespective of whether a segregated or coupled approach is being used to solve for the velocity and pressure fields. The advantage of this approach is that it is fast, as there no extra global steps. However, the significant disadvantage is that the actual pressure and velocity are affected. Empirically, we observe that this approach leads to fluctuations in the pressure field, and often also leads to unstable simulations. 

An alternative approach to use the artificial velocity divergence introduced in Eq.\,\eqref{Eq:DivVcorr}, which will be employed in the present work, is to solve for an artificial displacement independent of the main numerical scheme for velocity and pressure. For this, we solve a pressure Poisson equation similar to that obtained in classical projection schemes \cite{Brown2001, Chorin1968}
\begin{equation}
	\label{Eq:VolCorr}
	-\frac{1}{\rho} \frac{\partial \rho}{\partial p} \frac{1}{\Delta t} p^{\text{artificial}} + 
	\nabla \cdot \left( \Delta t \frac{1}{\rho} \nabla 
	p^{\text{artificial}}  \right) = 
	\left( \nabla \cdot \vec{v} \right) ^{\text{artificial}} \,,
\end{equation}
for the artificial pressure $p^{\text{artificial}}$. This is subsequently used to create the artificial displacement, which can be computed locally at each point
\begin{equation}
\label{Eq:PointDispArt}
	\Delta \vec{x}^{\,\text{artificial}} = - \Delta t \left( \frac{\Delta t}{\rho} \nabla p^{\text{artificial}} \right) \,.
\end{equation}

Since this artificial pressure is only used to create an artificial point displacement, the actual velocity and pressure are not affected. As a result, this approach does not introduce any spurious effects into the numerical solution.

\subsection{Volume correction summary} 
\label{sec:VolCorrAlgo}
The resultant procedure for a volume corrected meshfree collocation flow solver is summarized in Algorithm~\ref{alg:VolCorr}, where the italicized steps indicate those not done in a typical flow solver. 

\begin{algorithm}
    \caption{The volume correction mechanism} \label{alg:VolCorr}
    \begin{algorithmic}[1]
        \State \emph{Initialize representative masses} (Eq.\,\eqref{Eq:RepMassInitial})
        \While{Time-stepping loop}
            \State Move point cloud in a Lagrangian sense \cite{Suchde2018_PCM}
            \State Update neighbour tree
            \State Add new points, flag points to be deleted
            \State Volume computation (Section~\ref{sec:Volume})
            \State \emph{Update representative masses for addition, deletion} (Sections~\ref{sec:Add}, \ref{sec:Delete})
            \State Delete points flagged for deletion
            \State \emph{Smooth representative masses} (Section~\ref{sec:MassAdaptation})
            \State \emph{Compute Representative densities} (Eq.\,\eqref{Eq:RepDens})
            \State Flow solver: velocity, pressure solve, stress tensor update if applicable
            \State \emph{Compute artificial divergence} (Eq.\,\eqref{Eq:DivVcorr})
            \State \emph{Compute artificial pressure} (Eq.\,\eqref{Eq:VolCorr})
            \State \emph{Artificial point movement} (Eq.\,\eqref{Eq:PointDispArt})
            \State Post-processing calculations
        \EndWhile
    \end{algorithmic}
\end{algorithm}




\section{Numerical Examples}
\label{sec:Numerical}

We now present the application of the introduced representative mass and volume correction frameworks to a series of numerical tests of varying complexities, starting from simple academic test cases to actual industrial test cases. The test cases used and the relevance of each test case is summarized in Table~\ref{tab:Testcases}.

To quantify the loss in volume, we use an error defined as the relative deviation of the numerical volume of the fluid being simulation from the expected value. Thus, the error is given by
\begin{equation}
\label{Eq:Verr}
	\epsilon_V = \frac{|  V_{\text{analytical}} - V |}{ V_{\text{analytical}} }\,,
\end{equation}
where $V = \sum_{i=1}^N V_i$ is the sum of volumes of each point in the domain, and $V_{\text{analytical}}$ is the analytical volume expectation. 
\begin{table}
	\caption{Numerical experiments used to highlight the effectiveness of the representative mass and volume correction frameworks. 
 $d$ shows the dimensionality of the problem, with the computational domain $\Omega \subset \mathbb{R}^d$. For the three dimensional test cases, the Reynolds number is indicated by $Re$. }
	\centering
	\label{tab:Testcases}
	{  
    \begin{tabular}{|p{0.02\textwidth}|p{0.3\textwidth}|p{0.55\textwidth}|}
	\hline
    & Test case &  Comments \\ 
	\hline \hline
1& Adaptive refinement              &  $d = 2$ \newline 
                                        Simplified test for stationary fluid  \newline
                                        Illustration of representative mass adaptation  \\\hline                                        
2& Dam break                        &  $d = 2$ \newline
                                        Common meshfree benchmark test case 
\\\hline
3& Drops falling and collection     &  $d = 3$ \newline
                                        Convergence with spatial resolution \newline
                                        $Re \approx 15$ \\\hline
4& Partially filled screw conveyor  &  $d = 3$ \newline
                                        Industrially motivated test case \newline
                                        Fluid disappears without volume correction  \newline
                                        $Re \approx 0.2$ \\\hline                                        
5& Automotive water crossing        &  $d = 3$ \newline
                                        Industrial test case \newline
                                        Violent free surface flows  \newline
                                        $Re \approx 10^7$ \\\hline
	\end{tabular}}
\end{table}

\subsection*{Numerical scheme}

The representative mass and volume correction frameworks introduced in the present work can be used with any choice of time integration scheme, and any meshfree collocation method for derivative discretization. We briefly introduce the methods used in the numerical examples for the sake of completeness. 

We use a Generalized Finite Difference Method~(GFDM) approach to compute numerical derivatives. The GFDM is a strong form meshfree collocation approach that has been widely used \cite{Fan2018, Luo2016, Xia2021, Zhan2022}, and shown to be a robust framework for a variety of flow applications \cite{Moller2007, Tramecon2013} including flow through porous media \cite{Rao2022}, non-Newtonian flow \cite{Saucedo2021, Veltmaat2022}, and even for soil mechanics \cite{Michel2021}, and elasticity problems \cite{Saucedo2022}. The GFDM approach used here is second order accurate, with polynomial basis functions. 


The overall time integration procedure is explained in Algorithm~\ref{alg:VolCorr}. The basic scheme starts with Lagrangian movement with a second order approach \cite{Suchde2018_PCM}. This is followed by a monolithic scheme where the velocity and pressure are solved together in a single system, using an implicit first-order accurate time integration scheme. A verification and validation of the basic scheme used (without volume correction), and the implementation, can be found in our earlier work \cite{Drumm2008, Suchde2017}. For more details of the scheme used, we refer to \cite{Jefferies2015, Kuhnert2014, Suchde2018_INSE}. 
The numerical solvers used and the novel frameworks introduced in the present work are part of the in-house developed software suite MESHFREE \cite{MESHFREE}.

\subsection{Adaptive Point Cloud Refinement}

\begin{figure}
  \centering
  \includegraphics[width=0.27\textwidth]{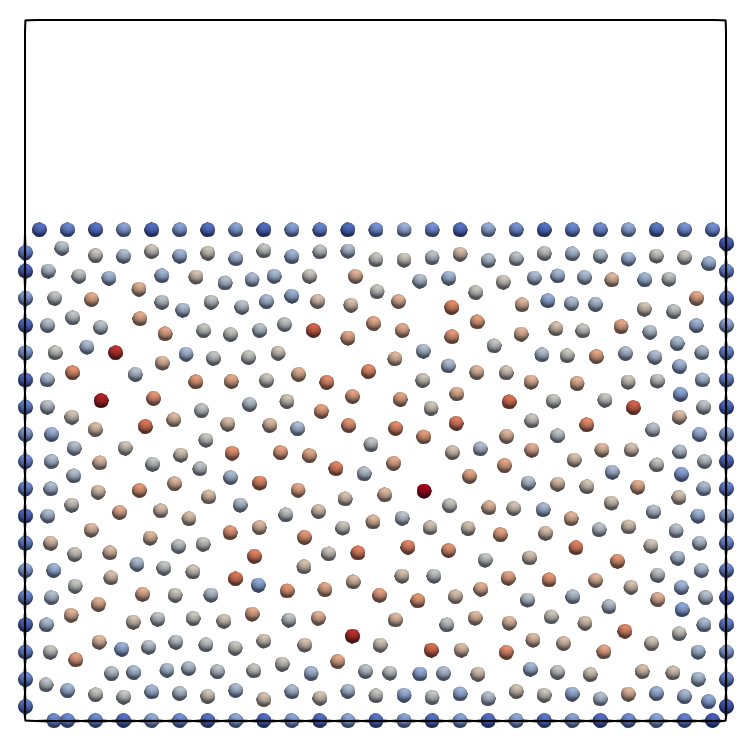}    
  \includegraphics[width=0.27\textwidth]{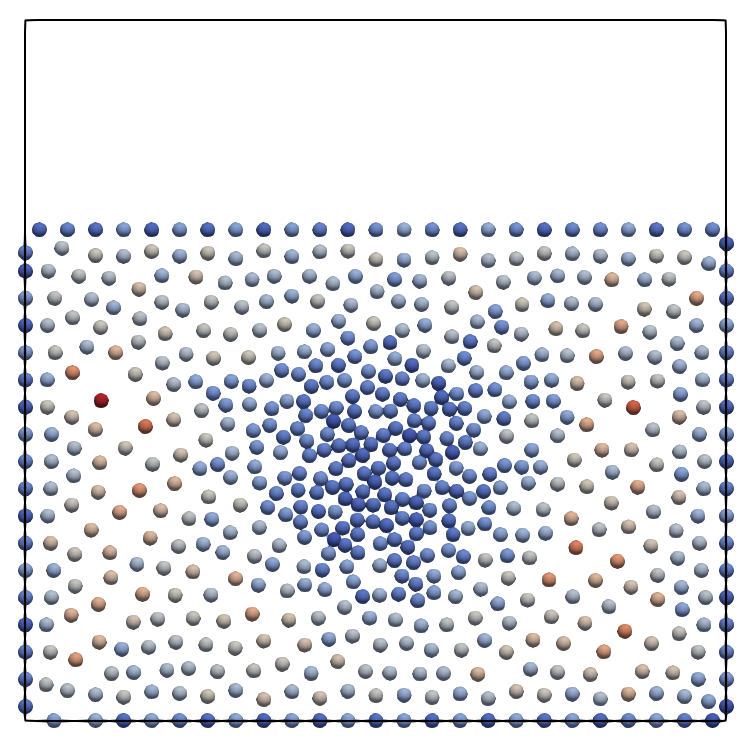} 
  \includegraphics[width=0.27\textwidth]{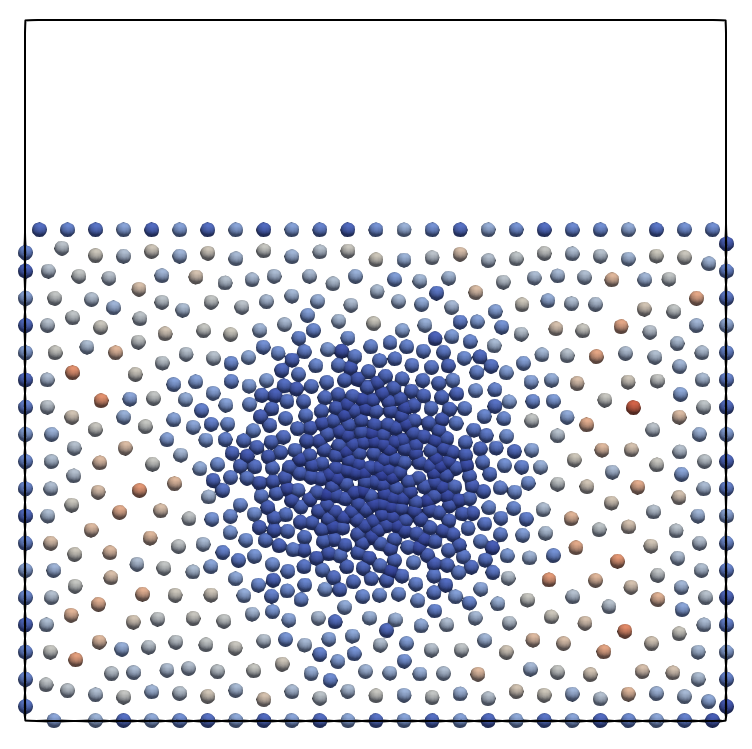}
  \includegraphics[width=0.12\textwidth]{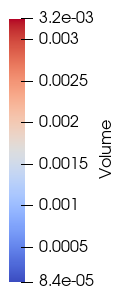}
	\caption{Adaptive refinement: Gradually refining the point cloud. The computational domain is marked in black. The stationary fluid only fills a part of the domain. The colour indicates the numerical volume of each point. The evolution of the numerical volume and the representative mass are shown in Figure~\ref{Fig:AdaptiveRefinement_MassVolume}. } 
  \label{Fig:AdaptiveRefinement}%
\end{figure}

For the first test case, we consider a stationary fluid with a free surface in a square container. In the middle of the domain, the point cloud is refined (see Figure~\ref{Fig:AdaptiveRefinement}) till the total number of points in the domain approximately doubles. The point cloud is then coarsened towards a uniform resolution, with the simulations terminated at a fixed time $t=2$. This example illustrates one of the main advantages of meshfree collocation methods over meshfree particle methods: the ease of performing adaptive refinement (see Section~\ref{sec:Meshfree}). Since no external force is applied, the fluid should remain at rest, which is also observed in the simulations. 

\begin{figure}
  \centering
  \includegraphics[width=0.44\textwidth]{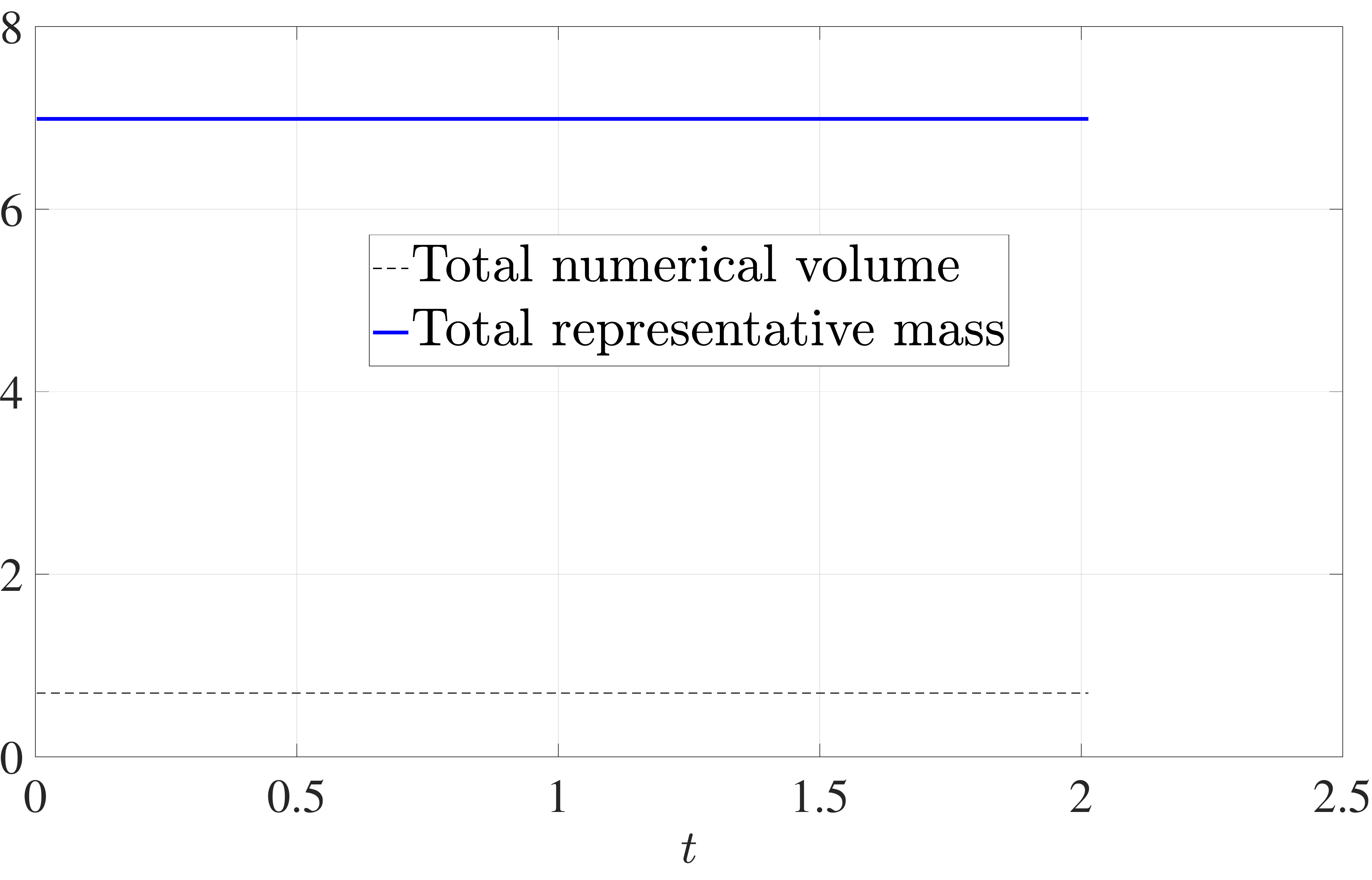}    
  \includegraphics[width=0.47\textwidth]{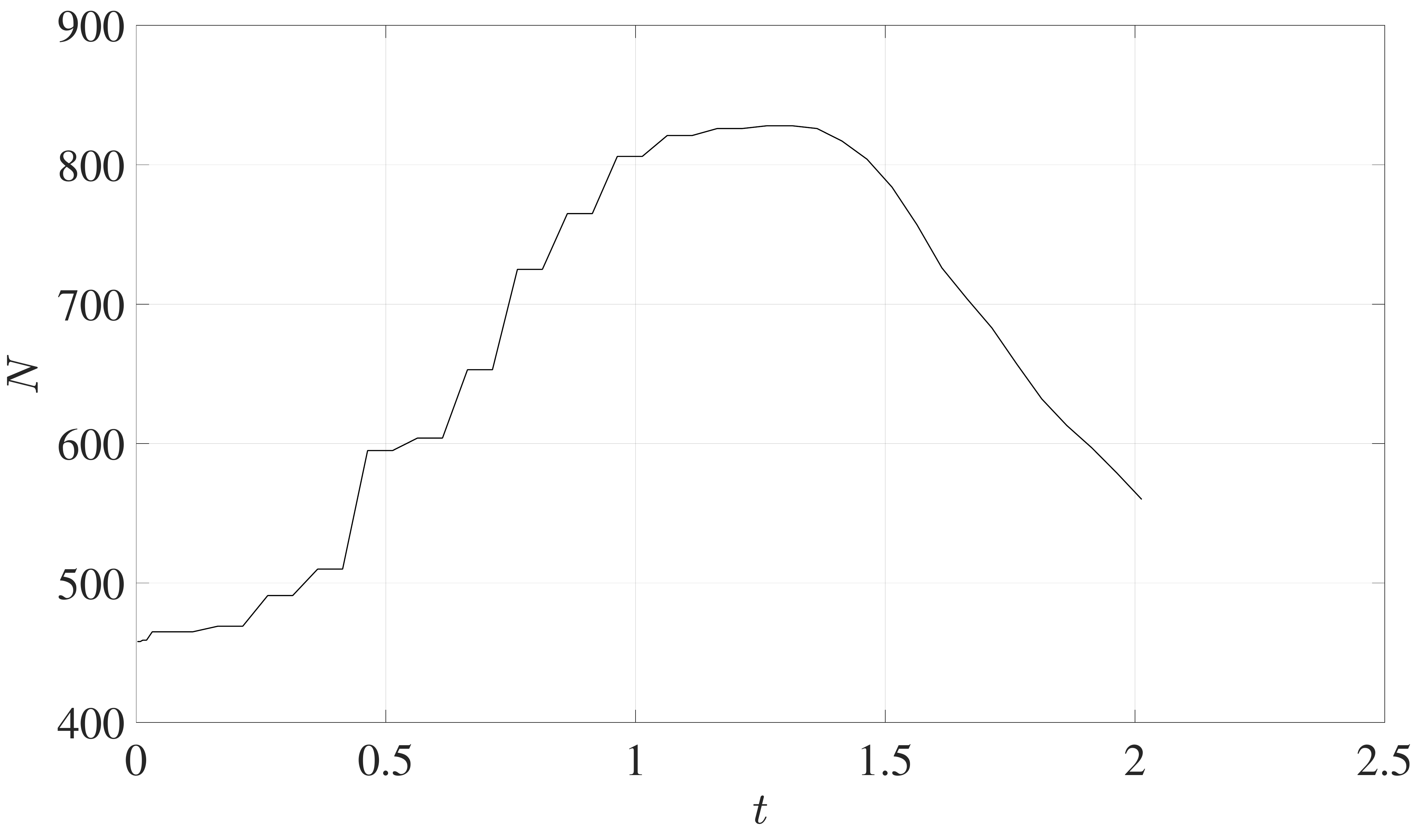} 
	\caption{Adaptive refinement test case: Evolution of the total mass ($\sum_{i=1}^N \hat{m}_i$) and volume ($\sum_{i=1}^N V_i$) in the computational domain as the point cloud is refined~(left), and the evolution of the number of the points ($N$) in the domain~(right). } 
  \label{Fig:AdaptiveRefinement_MassVolume}%
\end{figure}

The simulation domain is taken as a unit square $[0,1]^2$, with the free surface located at $0.7$ in the height direction. Thus, the analytical volume occupied by the domain is $0.7$. Figure~\ref{Fig:AdaptiveRefinement_MassVolume} shows that the total numerical volume $V = \sum_{i=1}^N V_i$ remains constant during both refinement and coarsening, and closely matches the analytical volume. Quantitatively, the total numerical volume deviates from the analytical one by less than $0.5 \%$. This shows that using the locally defined tessellations for determining volumes (see Section~\ref{sec:Volume}) produces very good and consistent results. It is important to note here that the presence of sharp boundaries in the present case, and curved boundaries in a general scenario, will always result in imperfections in the prescription of discrete volumes. 

Figure~\ref{Fig:AdaptiveRefinement_MassVolume} also shows that the total representative mass $\hat{\mathbb{M}} = \sum_{i=1}^N \hat{m}_i$ in the domain remains conserved during the adaptive refinement process. The fluid density is taken as $\rho = 10$, thus giving the analytical total mass as $7$ units. The numerical mass closely matches this value throughout the simulation, with the error bounded by the same amount as the deviation in the total volume.  

The examples illustrates that the representative masses adaptation works well while adding and deleting points. This also serves as a preliminary test to show that no spurious effects are introduced by the volume correction mechanism. 

%
%

%

\subsection{Dam Breaking}




We now consider a two dimensional dam break problem, which is a classical benchmark test case for free surface flows. 
The dam break simulation is run till the fluid comes to a rest. To prevent the influence of the numerical definition of volumes, the final volume of the fluid is determined by the height of the fluid at rest multiplied by the width of the domain. This is then compared to the analytical initial volume. 

\begin{figure}
  \centering
  \includegraphics[width=0.7\textwidth]{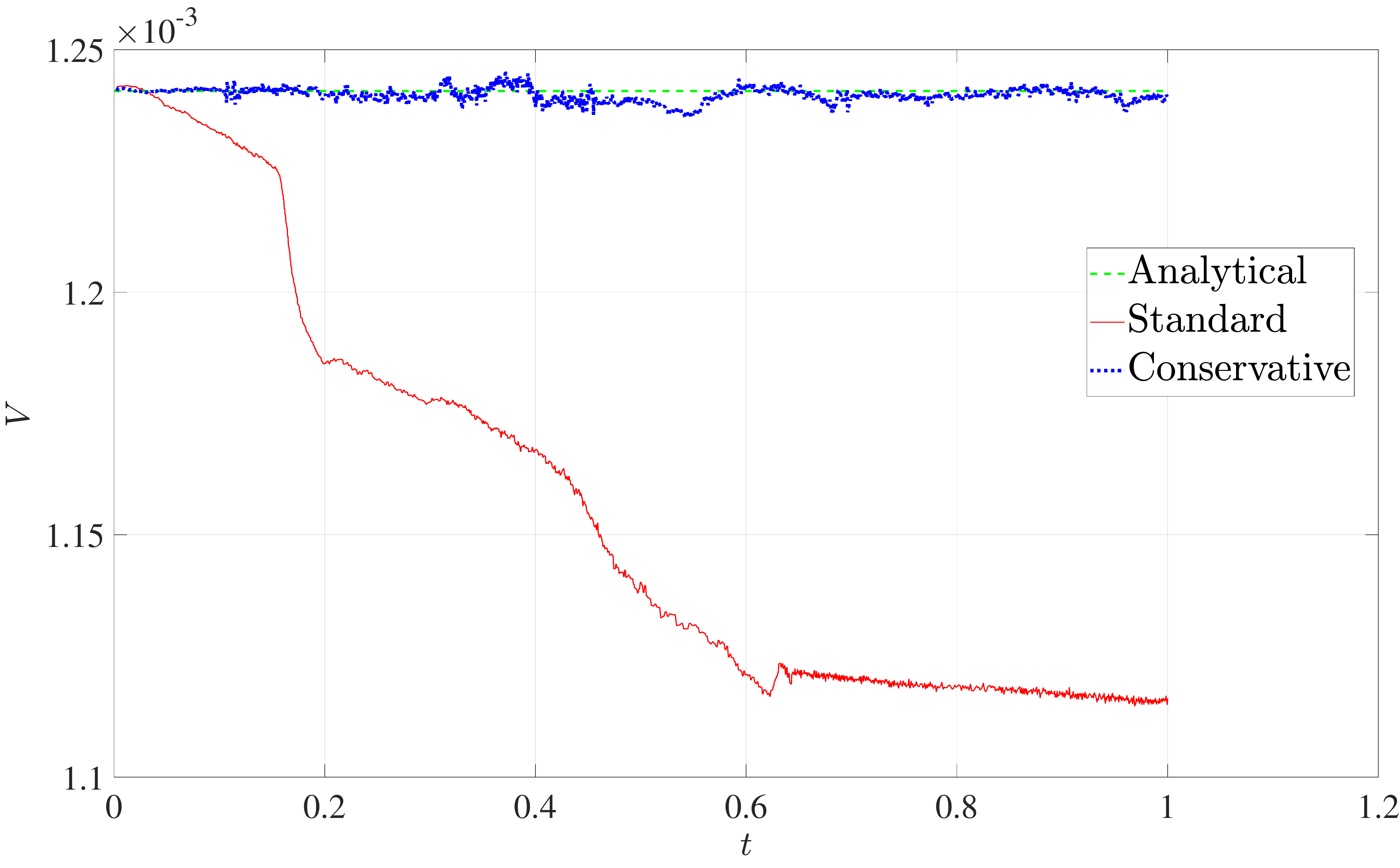}    
	\caption{Dam breaking test case: Evolution of numerical volume for the collocation simulations, with and without the volume conservation algorithm.} 
  \label{Fig:DamBreak_Vol}%
\end{figure}

The evolution of the numerical volume for the standard collocation approach and the volume conservative method is shown in Figure~\ref{Fig:DamBreak_Vol}. The standard approach has a steady drop in numerical volume till the fluid starts coming to a rest, with the biggest volume drop occurring when the fluid reaches the wall. 
On the other hand, the volume conservative simulations only show minor fluctuations of volume around the analytical volume. 

\subsection{Droplets Falling}

\begin{figure}
  \centering
  \subfloat[(a)]{\includegraphics[width=0.4\textwidth]{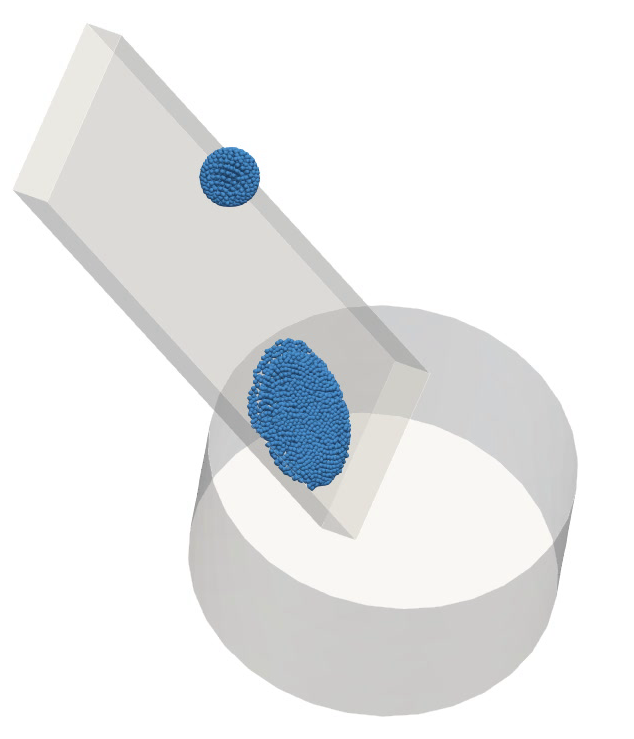}    }
  \subfloat[(b)]{\includegraphics[width=0.4\textwidth]{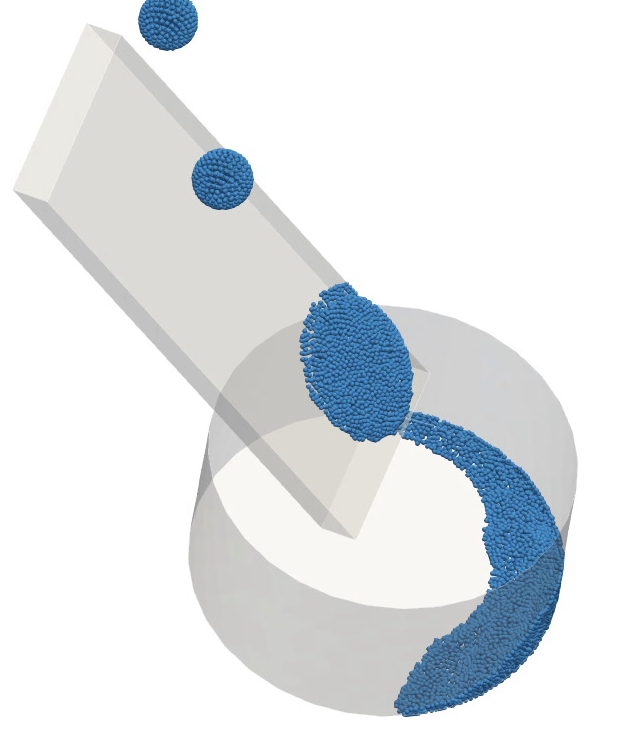}}\\  
  \subfloat[(d)]{\includegraphics[width=0.4\textwidth]{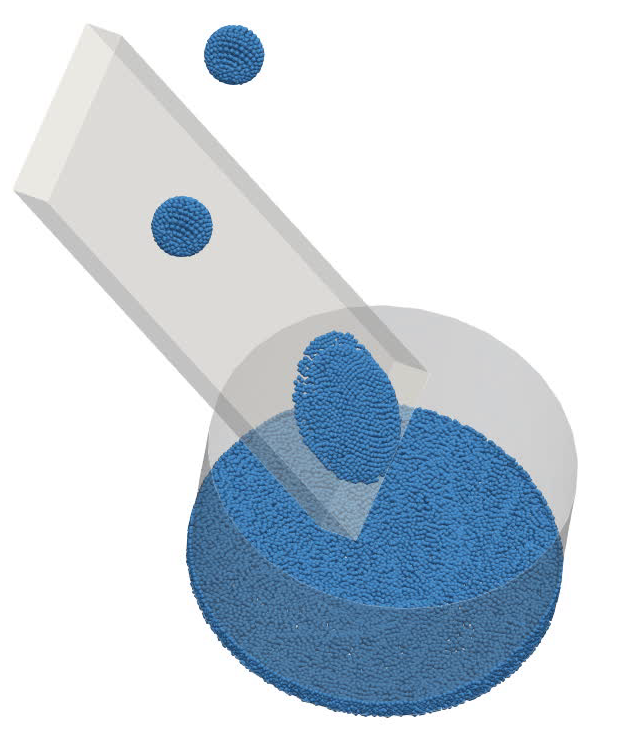}}
  \subfloat[(c)]{\includegraphics[width=0.4\textwidth]{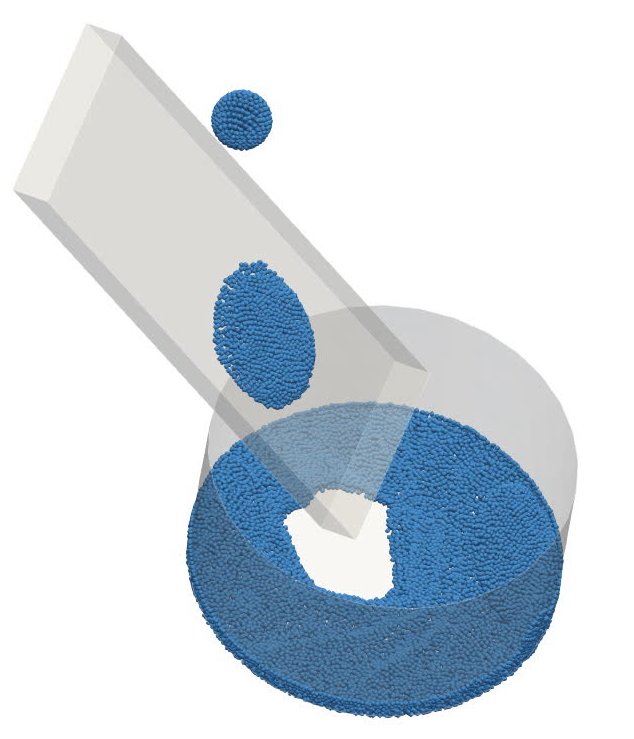}}
	\caption{Droplets falling test case: Volumetrically resolved fluid droplets falling onto an inclined plate and then collecting in a cylindrical container. Image sequence is clockwise from top-left. } 
  \label{Fig:DropletsFalling}%
\end{figure}

While the earlier two test cases had two dimensional spatial domains, henceforth, all examples considered are in $\mathbb{R}^3$. We now consider a more complex case of rain-like droplets falling onto an inclined plate and collecting in a bowl below the plate, as shown in Figure~\ref{Fig:DropletsFalling}. Upon impinging on the inclined plate, the droplets slide down and then collect in a cylindrical container below. Each droplet considered is volumetrically resolved. The inflow of droplets of a constant diameter occurs above the inclined plate at a fixed rate. The height of the droplets above the inclined plate, and viscosity of the fluid considered is chosen such that the droplet does not splash upon impact. A test case with splashing and spraying fluid is considered later in Section~\ref{sec:WaterCrossing}.  
%

Each fluid droplet is spherical with diameter $0.15 \text{ m}$. A newly generated droplet has no initial velocity, and falls due to gravity. The fluid considered has density $\rho = 1 \text{ kg/m}^3$, and dynamic viscosity $\eta = 0.051 \text{ Pa s}$. This results in a Reynolds number of about $Re \approx 15$. 
Surface tension and contact angle hysteresis effects are not considered. Homogeneous Neumann pressure boundary conditions are applied on all wall boundaries. Slip boundary conditions are applied for the velocity at both the inclined plate and the container. 

Since fluid enters the domain at a fixed rate at the inflow, the total geometrical volume of the fluid domain considered is known analytically at every time step,  $V_{\text{analytical}} = V_{\text{drop}} \sum_{\text{drops}} 1$, where $V_{\text{drop}}$ is the volume of single droplet, and the summation is over the number of droplets injected into the domain until the current time. Thus, the analytical volume expectation has a jump every time a new droplet is added to the simulation domain.

\begin{figure}
  \centering
  \subfloat[(a) $h = 0.02$. Points per droplet $= 3\,807$ ]{\includegraphics[width=0.7\textwidth]{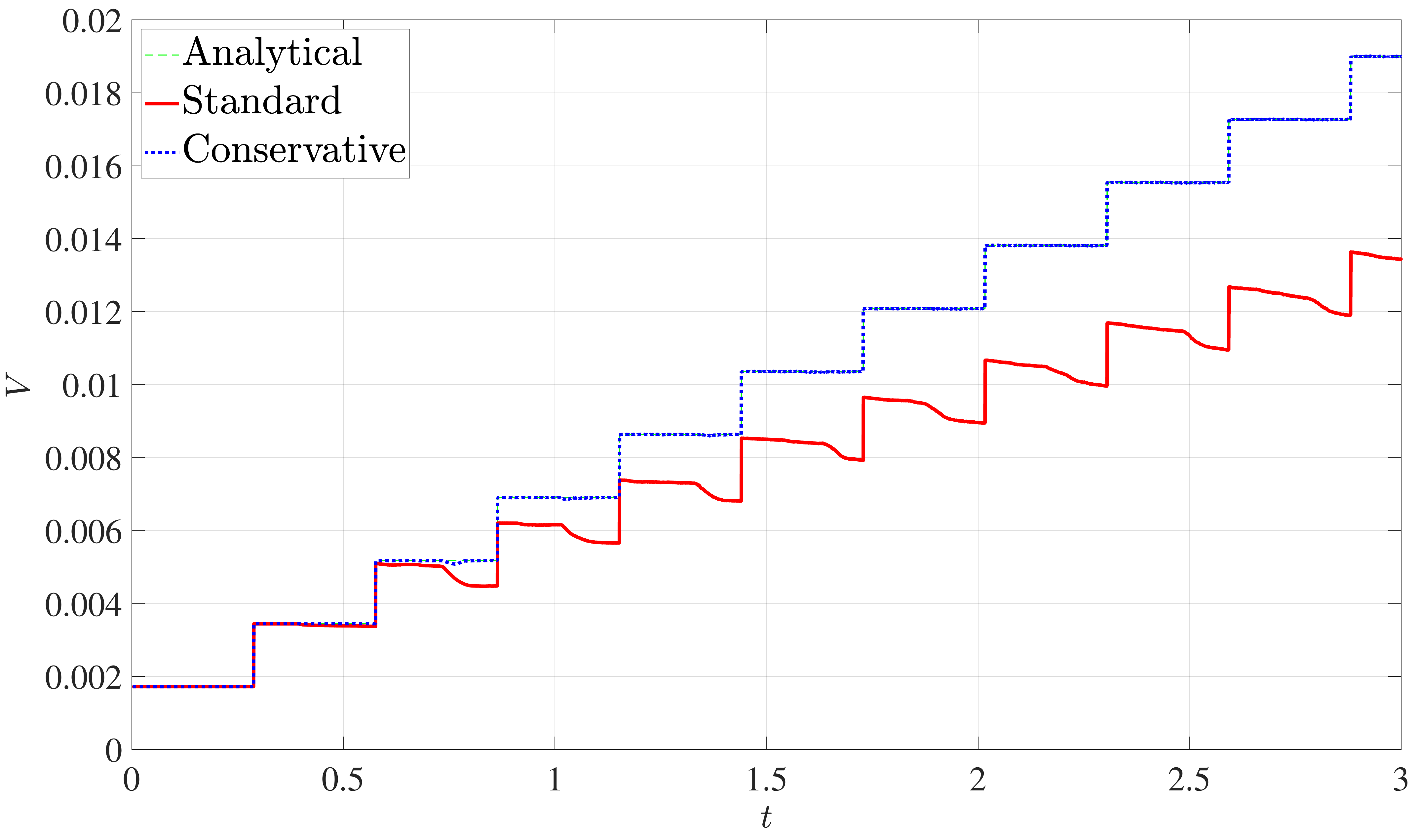} }\\
  \subfloat[(b) $h = 0.04$. Points per droplet $= 29\,353$]{\includegraphics[width=0.7\textwidth]{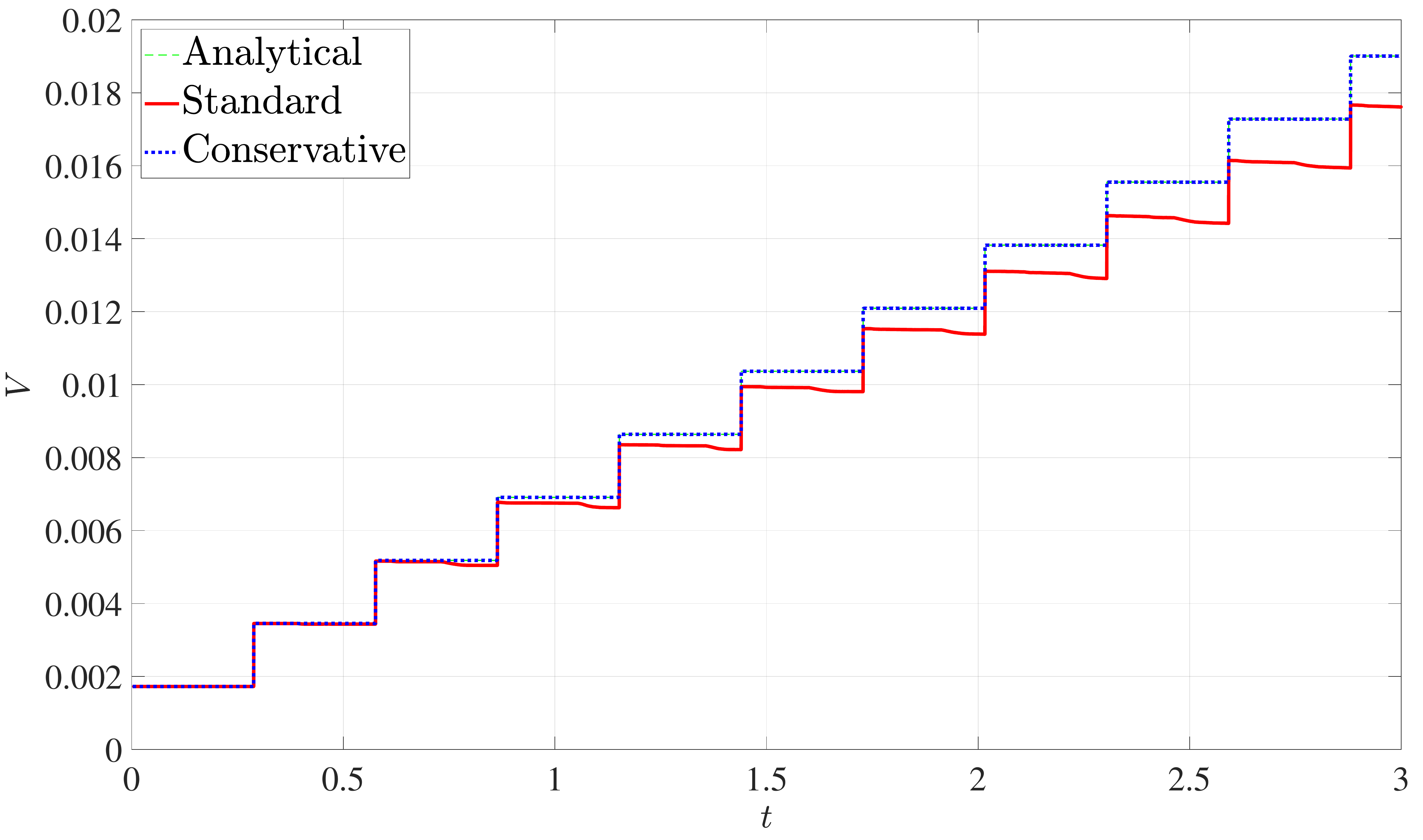} }
	\caption{Droplets falling test case: Numerical volume compared to analytical volume. In both figures, the actual analytical volume (dashed green line) and the numerical volume in the conservative simulations (dotted blue line) are almost overlapping (see Table~\ref{tab:DropFallTab} for a quantified comparison), while the numerical volume in the standard simulations (solid red line) deviates significantly.} 
  \label{Fig:DropletsFalling_Vol}%
\end{figure}

The total numerical volume in the domain $V$ as compared to the analytical expectation is plotted in Figure~\ref{Fig:DropletsFalling_Vol}. The figure shows the evolution of the numerical volume for both the standard simulations and the conservative one, for two different resolutions. In both cases, the conservative simulation shows a numerical volume very close to the expected one. On the other hand, the standard simulation deviates significantly from the analytical expectation. The biggest drop in volume occurs each time a droplet impinges on the inclined plate, with a further decline in numerical volume after it reaches the container. 

A quantification of errors in volumes (see Eq.\,\eqref{Eq:Verr}) for both the standard and conservative simulations is tabulated in Table~\ref{tab:DropFallTab} for multiple resolutions. The volume error for the conservative simulations is almost constant as the resolution is made smaller. For the coarsest simulation considered, $h = 0.04$ which corresponds to $619$ points per droplet, the volume error for the standard case is $29 \%$, while that for the conservative case is three orders of magnitude smaller at $0.047 \%$. Whereas for the finest simulation considered, $h = 0.01$ which corresponds to $29353$ points per droplet,the error in the conservative case is two orders of magnitude smaller than that in the standard case. The number of points needed to resolve each droplet for the different resolutions used is listed in Table~\ref{tab:DropFallTab}, along with the total number of points in the simulation domain at the end of the simulation at $t = 3$. Note that the number of point at the end of the simulation varies slightly between the standard and conservative simulations due to the addition and deletion algorithms (see Sections \ref{sec:Meshfree} and \ref{sec:Notation}).
\begin{table}
	\caption{Droplets falling test case: Error in volume, $\epsilon_V$ (see Eq.\,\eqref{Eq:Verr} ) at $t=3$, number of points per droplet, and total number of points in the simulation domain at the end of the simulation ($t=3$) for the different resolution sizes considered.}
	\centering
	\label{tab:DropFallTab}
	{  
    \begin{tabular}{|c|c|c|c|c|}
	\hline
    Resolution   & Points per droplet &  Points at $t=3$  & $\epsilon_V$ standard &  $\epsilon_V$ conservative  \\ 
	\hline \hline
$h = 0.04$   & $\phantom{00}\,619$   &  $3.1 \times 10^4$  & $2.9 \times 10^{-1}$ &   $4.7 \times 10^{-4}$  \\\hline
$h = 0.02$   & $\phantom{0}3\,807$   &  $1.8 \times 10^5$  & $7.3 \times 10^{-2}$ &   $5.1 \times 10^{-4}$  \\\hline
$h = 0.01$   & $29\,353$             &  $6.8 \times 10^5$  & $2.6 \times 10^{-2}$ &   $4.8 \times 10^{-4}$  \\\hline%
	\end{tabular}}
\end{table}
%

%
%
%
%
%
%
%
%


\subsection{Partially Filled Screw Conveyor}

Screw conveyors are used for material processing, for example for polymer extrusion. They use sophisticated systems with interacting screws and often complex rheology is involved. Some systems are only partially filled with material thus giving rise to a free surface between the material and surrounding atmosphere. Meshfree methods come with a great advantage when simulating such partially filled systems. However, volume loss can be a severe issue especially at machine parts which are moving fast against each other like a screw and its casing. This is a problem often seen in low resolution meshfree simulations. This problem could be circumvented with very small spatial and temporal resolutions. However, that would effectively make simulations of larger systems impossible. Using the proposed volume correction algorithm enables us to simulate large systems with moderate resolution settings.

\begin{figure}
  \centering
  \includegraphics[width=0.95\textwidth]{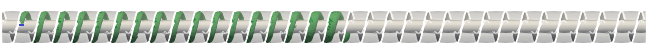}      
	\caption{Screw conveyor test case: Illustration of the domain. The fluid is marked in green, and the fluid inlet is marked in blue on the left of the domain. The screw rotates clockwise around its axis, while the casing is fixed. As the screw rotates, the fluid is transported from the left to the right side in the figure. } 
  \label{Fig:ExtrusionGeometry}%
\end{figure}

Motivated by such an industrial polymer extrusion problem, we demonstrate the effects of volume correction using a simplified test case of a screw conveyor as shown in Figure \ref{Fig:ExtrusionGeometry}. The screw is rotating clockwise around the x axis (pointing from left to right in the figure). The screw is tightly enclosed in a non-rotating casing. Polymer melt is injected on the left. The inlet is modeled as a small circular surface which rotates together with the screw. The material is then transported towards the right side by the moving screw. For simplicity, the material is modeled as a Newtonian fluid with constant viscosity. For a real setting, more complex rheological models are needed. The fluid used has a desity $\rho = 10^3 \text{ kg/m}^3$, and dynamic viscosity $\eta = 200 \text{ Pa s}$. The diameter of the screw is $0.04 \text{ m}$, which results in a Reynolds number of about $Re \approx 0.2$.

\begin{figure}
  \centering
  \includegraphics[width=0.95\textwidth]{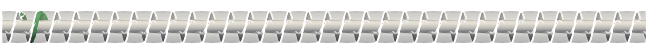}\\
  \includegraphics[width=0.95\textwidth]{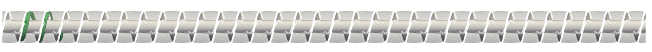}\\
  \includegraphics[width=0.95\textwidth]{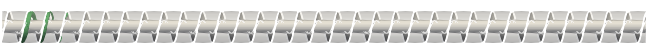}\\
  \includegraphics[width=0.95\textwidth]{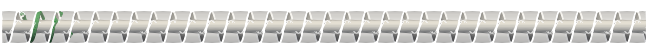}\\
  \includegraphics[width=0.95\textwidth]{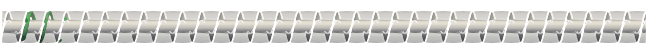}\\
	\caption{Screw conveyor test case: Evolution of extrusion without the volume correction mechanism. The fluid is marked in green. This highlights a common problem in meshfree methods where the fluid volume disappears when the resolution used is not fine enough. In contrast, this problem is no longer present in the volume corrected simulations, as shown in Figure~\ref{Fig:Extrusion_localVC}. } 
  \label{Fig:Extrusion_noVC}%
\end{figure}
\begin{figure}
  \centering
  \includegraphics[width=0.95\textwidth]{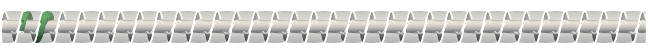}\\
  \includegraphics[width=0.95\textwidth]{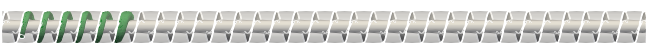}\\
  \includegraphics[width=0.95\textwidth]{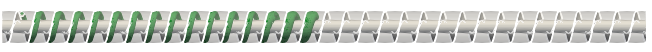}\\
  \includegraphics[width=0.95\textwidth]{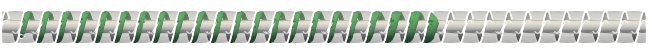}\\
  \includegraphics[width=0.95\textwidth]{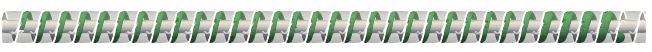}\\
	\caption{Screw conveyor test case: Evolution of extrusion without the volume correction mechanism. The fluid is marked in green. The issue of mass loss present in the standard simulations (Figure~\ref{Fig:Extrusion_noVC}) is no longer present here. } 
  \label{Fig:Extrusion_localVC}%
\end{figure}

The evolution of the extrusion process is shown in Figure~\ref{Fig:Extrusion_noVC} without the volume correction mechanism. Mass is lost quickly, mainly at the edge between screw and casing which are moving against each other. The loss happens so fast that after two or three revolutions no material is left. The material never reaches the right side of the device. In order to have material reach the right side for the setting without volume correction the resolution would have to be chosen extremely small. The same setting with volume correction enabled is shown in Figure~\ref{Fig:Extrusion_localVC}. Both simulations use the same resolution. No mass loss is observed for this setting. The material is transported by the screw and it reaches the right side after some time.


\subsection{Industrial Application: Automotive Water Crossing}
\label{sec:WaterCrossing}

\begin{figure}
  \centering
  \includegraphics[width=0.32\textwidth]{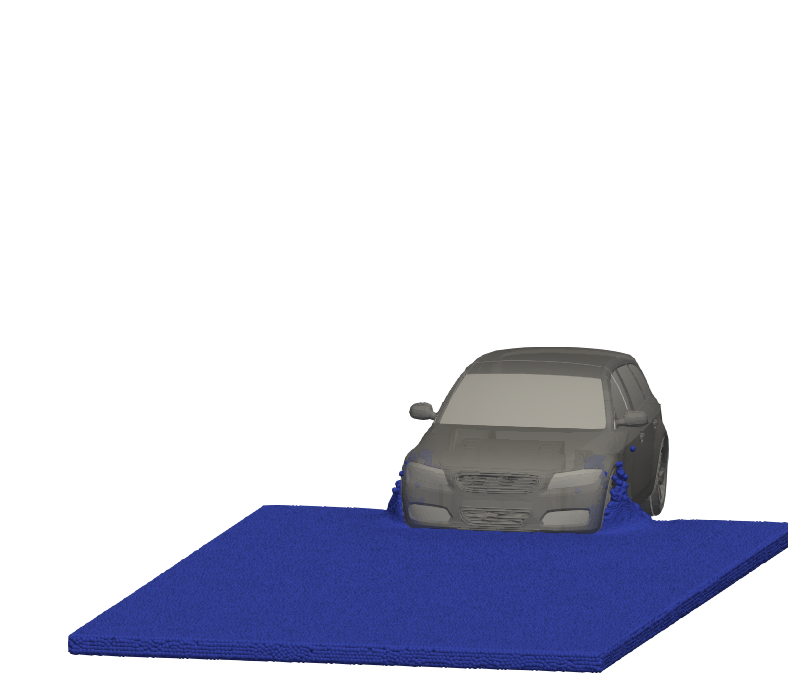}    
  \includegraphics[width=0.32\textwidth]{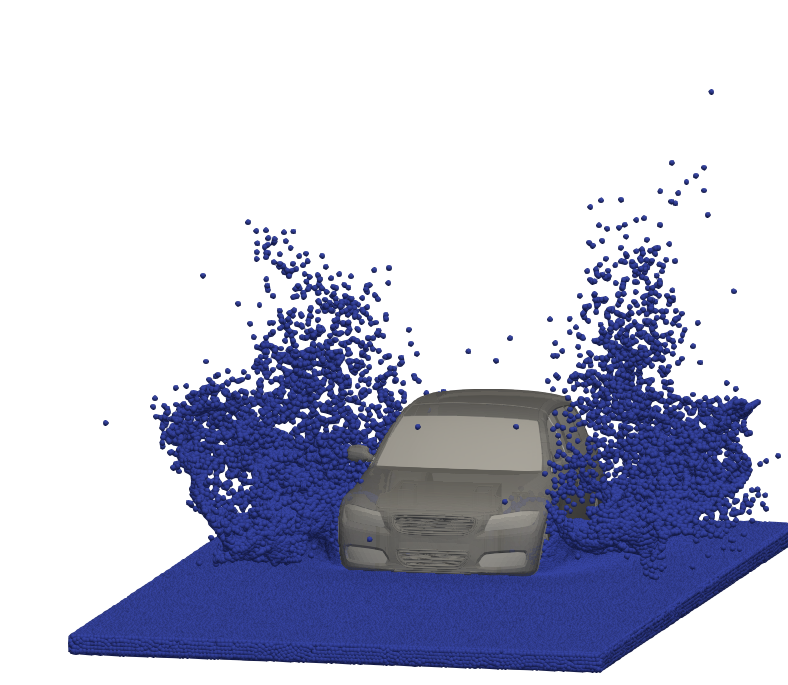}
  \includegraphics[width=0.32\textwidth]{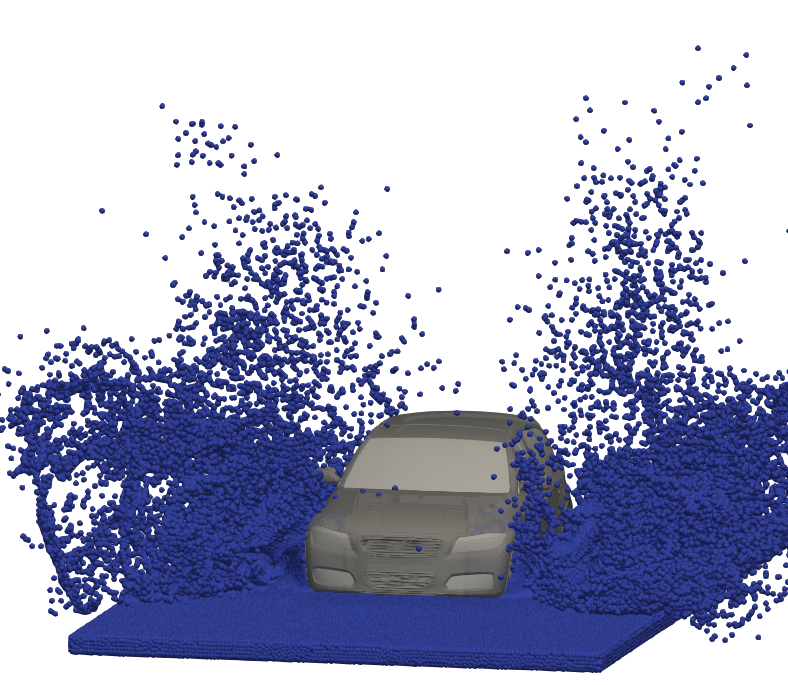}
	\caption{Automotive water crossing: A car crossing a shallow pool of water a constant velocity. The car is shown with a slight transparency to improve the visualization of the fluid.} 
  \label{Fig:WaterCrossing}%
\end{figure}

We now consider an actual industrial application of a car moving through a shallow pool of water, as shown in Figure~\ref{Fig:WaterCrossing}. Such simulations have been referred to as water crossing or water wading simulations. The primary goal in these simulations is typically to check how much water reaches different parts of the car, especially in the under-body of the car. An example of this is to determine how much water, if any, enters the air intake where no water should go. This test cases represents a much more complex geometry and a non-smooth flow pattern than that considered in previous sections. 

This test case is a perfect example of the need for meshfree methods over mesh-based ones. Meshing the complex under-body of the car can be quite challenging. Moreover, since the car is moving through the water, the mesh needs to be updated at every time step. The domain occupied by the fluid spraying is much larger than the initial domain of the pool, as can be seen in Figure~\ref{Fig:WaterCrossing}. Rather than meshing the entire domain, a meshfree approach only tracks points or particles where the fluid is actually present, with no discretization away from the fluid. Furthermore, the fluid sprays to regions where very small amount of fluid is present, which would require a very fine mesh to capture accurately. 

The ease of adaptive refinement near the car geometry in meshfree collocation approaches give them an advantage over particle-type meshfree methods (see Section~\ref{sec:Meshfree}). However, due the absence of a native notion of mass, it is not possible to determine how much water truly goes into different parts of the under-body. The introduction of the notion of representative masses in the present work solves this problem. 


The car geometry used is the open-source DrivAer car model \cite{Heft2012} \footnotemark. \footnotetext{The estate back configuration with the detailed under body is used.} 
The majority of the geometry is treated as a rigid body moving at a fixed velocity. The wheels and tyres rotate at a fixed angular velocity around their respective axle such that it matches with the linear velocity of the car. 
Since the application is intended towards cars crossing rain water, the fluid simulated is water with $\rho = 10^3 \text{ kg/m}^3$ and $\eta = 10^{-3} \text{ Pa s}$. This results in a Reynolds number of about $Re \approx 2 \times 10^7$. 



\begin{figure}
  \centering
  \includegraphics[width=0.6\textwidth]{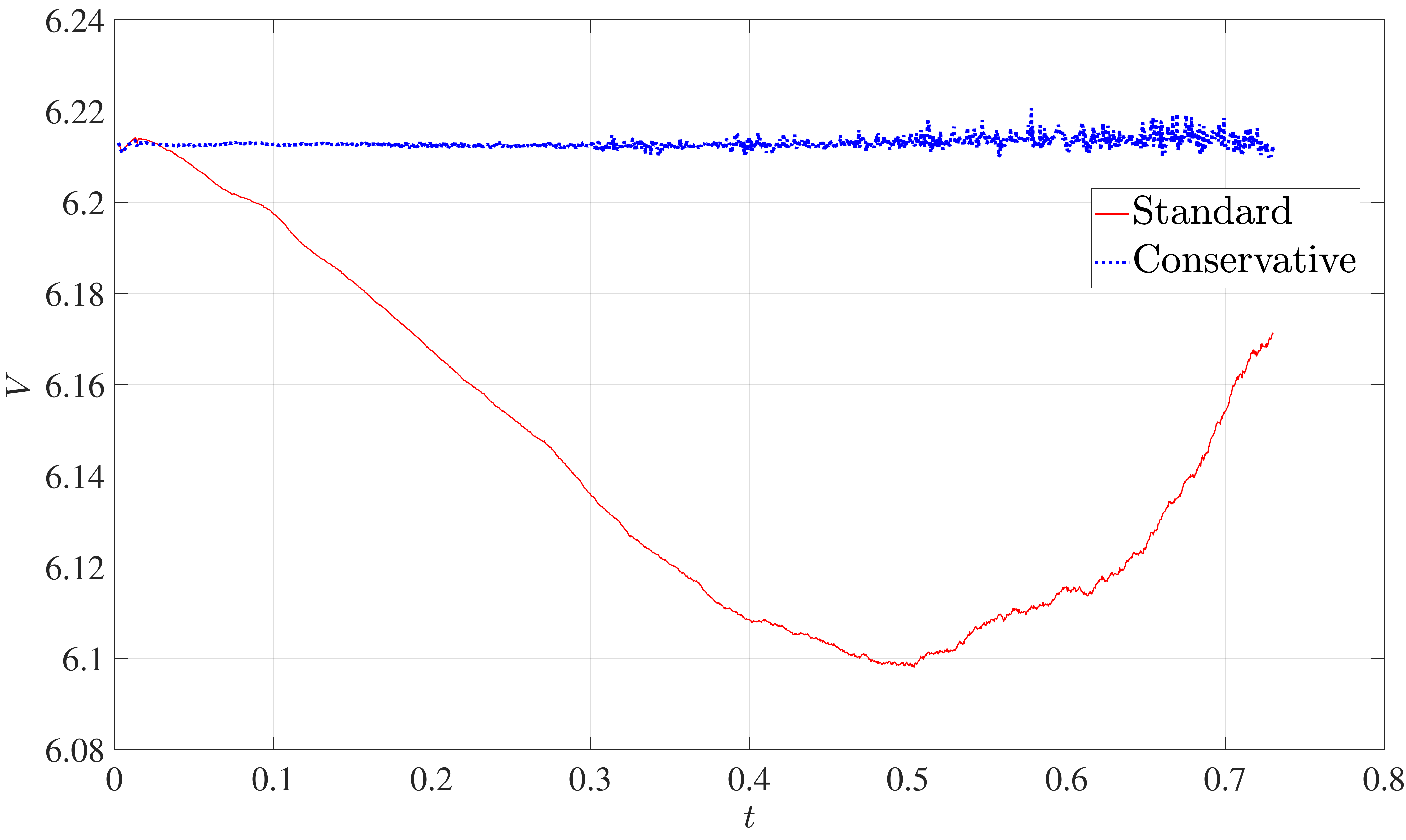}    
	\caption{Automotive water crossing: The total numerical volume of all points in the simulation for the conservative simulations (dotted blue line), and the standard simulations (solid red line). Since there is no inlet or outlet of fluid in the simulations, the volume should remain constant. In the conservative simulations, the volume is approximately constant, while it drops and then increases in the standard simulation. } 
  \label{Fig:WaterCrossing_Volume}%
\end{figure}

The evolution of the total fluid volume as the car crosses the water pool, for both the standard and conservative simulations, is shown in Figure~\ref{Fig:WaterCrossing_Volume}. Similar to the previous test cases, the numerical volume in the standard simulation drops as the simulation progresses, with the volume starting to increase once the water spray starts increasing. On the other hand, the volume in the conservative simulation remain approximately constant, with the fluctuations increasing once the fluid spray increases. 
About a $2 \%$ deviation in volume is observed in the standard simulations, despite the use of fine resolution in which the initial water pool contains about $2 \times 10^5$ points. 
%



\begin{figure}
  \centering
  \subfloat[Side view]{\includegraphics[width=0.49\textwidth]{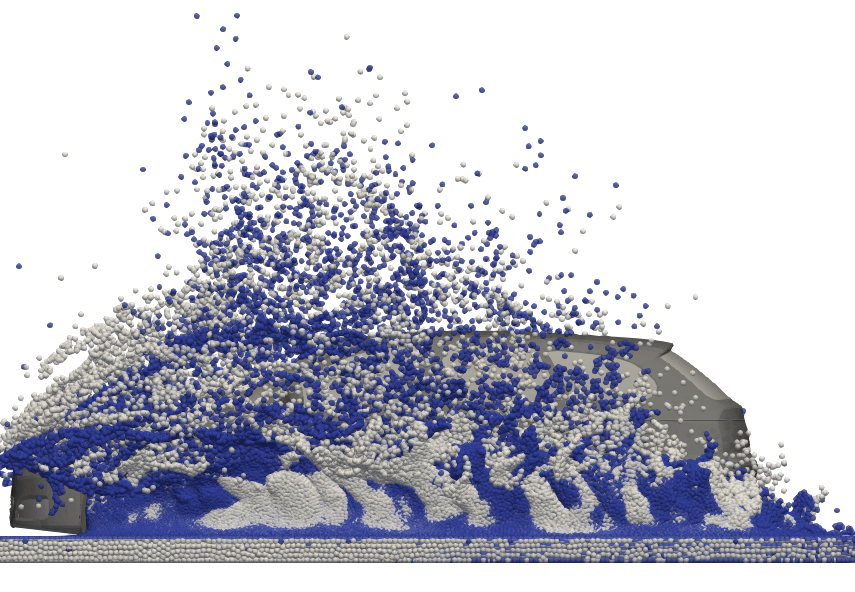}    }
  \subfloat[Bottom view with the water in the pool hidden]{\includegraphics[width=0.49\textwidth]{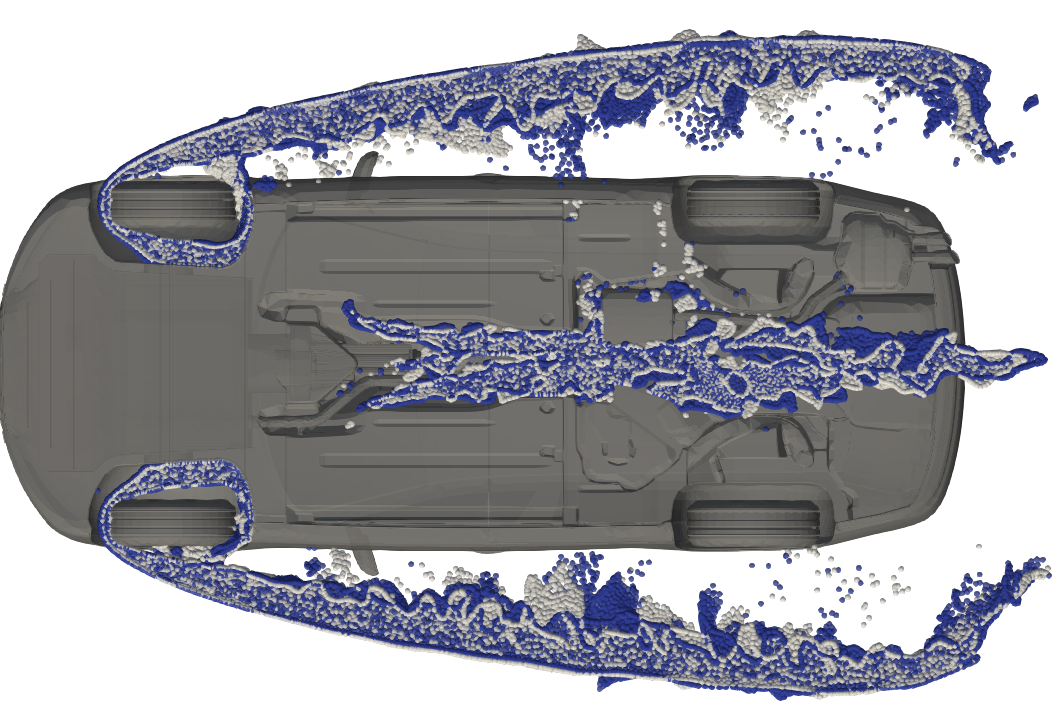}}    
	\caption{Automotive water crossing: Overlap of the standard simulation (white) and the conservative simulations (blue). A slight transparency is applied to both simulation results in both figures to enhance visualization. The figure shows that the standard and conservative simulations show very similar flow profiles.} 
  \label{Fig:WaterCrossing_Comparison}%
\end{figure}

A visual comparison of the spray of water in both cases in shown in Figure~\ref{Fig:WaterCrossing_Comparison}. The figure illustrates that the free surface spray patterns are similar in both cases, with slight differences only towards the tail of the spray. This illustrates that the actual difference made in the fluid profile is not significant. 

%

%

\section{Conclusion}
\label{sec:Conclusion}

In this work, we introduced a notion of mass, called representative mass, for points in meshfree collocation type methods. This was shown to be essential for proper post-processing, since a concept of mass is not inherently present in these methods. This work also presented a first overview of different kinds of conservation errors across different meshfree methods. We focused on the issue of volume conservation, for which we highlighted the role of the discrete definition of volume itself, while also giving an overview of different volume definitions in meshfree methods.  

Using the introduced representative masses, we computed a representative density. Based on the difference between the representative and actual densities, a novel volume correction mechanism was presented. Volume conservation was acieved by introducing an artificial velocity divergence like term to prescribe an artificial expansion or compression of the numerical domain. Our numerical results across multiple test cases of varying complexity show that the use of the volume conservation algorithm significantly improvements conservative behaviour, without affecting the flow patterns in any other way. We also showed that the introduced algorithm is very relevant for actual industrial flow problems as well.




While the method presented here was in the context of collocation type meshfree methods, it could be easily extended to be applied for particle type meshfree methods like SPH. Here, the artificial movement was done based on the relative difference between the representative density and the actual one. For particle type meshfree methods, the volume correction could be prescribed either in the same way, or based on the relative difference between the representative mass and actual mass.

\section*{Acknowledgements}

Pratik Suchde would like to acknowledge support from  the  European  Union’s  Horizon  2020
 research  and  innovation programme under the Marie Skłodowska-Curie Actions grant agreement No. 892761. 
Pratik Suchde and St\'ephane P.A. Bordas would like to acknowledge funding from the Institute of Advanced Studies, University of Luxembourg, under the AUDACITY programme. Pratik Suchde would also to thank Daniel Louis Louw for helpful discussions on FVM VoF, and Mohsen Abdolazadeh for helpful discussions on SPH. 


\bibliographystyle{abbrv}
\bibliography{./Conservation}

\end{document}